\documentclass[10pt,journal,twocolumn]{IEEEtran}
\usepackage{wrapfig}
\usepackage[]{graphicx}
\usepackage{booktabs}
\usepackage{fancyhdr}
\usepackage{amsmath}
\usepackage{amsfonts}
\usepackage{cite}
\usepackage{bm}
\usepackage{amssymb}
\usepackage{amsthm}
\usepackage{url}
\usepackage{multirow}
\usepackage{times}
\usepackage{enumitem}
\usepackage{comment}
\usepackage[linesnumbered,ruled]{algorithm2e}
\usepackage[colorlinks=true, linkcolor=black, citecolor=blue, urlcolor=blue]{hyperref}
\usepackage{siunitx}
\usepackage{mathtools}
\usepackage{balance}

\usepackage{tikz}
\usetikzlibrary{angles}
\usetikzlibrary{calc}
\usetikzlibrary{quotes}
\usetikzlibrary{fit}
\usetikzlibrary{positioning}
\usetikzlibrary{shadows}

\makeatletter
\let\MYcaption\@makecaption
\makeatother

\usepackage[font=footnotesize]{subcaption}

\makeatletter
\let\@makecaption\MYcaption
\makeatother

\newcommand{\brm}[1]{\bm{\mathrm{#1}}}
\newcommand{\abs}[1]{\left|{#1}\right|}
\newcommand{\round}[1]{\ensuremath{\lfloor#1\rceil}}

\DeclareMathOperator*{\argmin}{\arg\!\min}
\DeclareMathOperator*{\argmax}{\arg\!\max}

\let\originalleft\left
\let\originalright\right
\renewcommand{\left}{\mathopen{}\mathclose\bgroup\originalleft}
\renewcommand{\right}{\aftergroup\egroup\originalright}

\begin{document}

\title{All-Weather sub-50-cm Radar-Inertial Positioning}


\author{Lakshay~Narula,~\IEEEmembership{Student Member,~IEEE,}
        Peter~A.~Iannucci,~\IEEEmembership{Member,~IEEE,}
        and~Todd~E.~Humphreys,~\IEEEmembership{Member,~IEEE}
\thanks{L. Narula is with the Department of Electrical and Computer
Engineering, The University of Texas at Austin, Austin, TX, 78712 USA e-mail:
lakshay.narula@utexas.edu.}
\thanks{P.A. Iannucci and T.E. Humphreys are with the Department of Aerospace
Engineering \& Engineering Mechanics, The University of Texas at Austin,
Austin, TX, 78712 USA.}
}

\maketitle

\begin{abstract}
  Deployment of automated ground vehicles beyond the confines of sunny and dry
  climes will require sub-lane-level positioning techniques based on radio
  waves rather than near-visible-light radiation. Like human sight, lidar and
  cameras perform poorly in low-visibility conditions. This paper develops and
  demonstrates a novel technique for robust sub-50-cm-accurate urban ground
  vehicle positioning based on all-weather sensors.  The technique incorporates
  a computationally-efficient globally-optimal radar scan batch registration
  algorithm into a larger estimation pipeline that fuses data from
  commercially-available low-cost automotive radars, low-cost inertial sensors,
  vehicle motion constraints, and, when available, precise GNSS measurements.
  Performance is evaluated on an extensive and realistic urban data set.
  Comparison against ground truth shows that during \SI{60}{\minute} of
  GNSS-denied driving in the urban center of Austin, TX, the technique
  maintains $95^{\rm th}$-percentile errors below \SI{50}{\centi\meter} in
  horizontal position and \ang{0.5} in heading.
\end{abstract}


\newif\ifpreprint
\preprintfalse
\preprinttrue

\ifpreprint

\pagestyle{plain}
\thispagestyle{fancy}  
\fancyhf{} 
\renewcommand{\headrulewidth}{0pt}
\lfoot{\footnotesize \bf Copyright \copyright~2020 by Lakshay Narula, \\
Peter A. Iannucci, and Todd E. Humphreys}
\rfoot{\footnotesize \bf Preprint of the manuscript submitted for review}

\else

\thispagestyle{empty}
\pagestyle{empty}

\fi

\section{Introduction}
\label{sec:introduction}

\IEEEPARstart{D}{evelopment} of automated ground vehicles (AGVs) has spurred
research in lane-keeping assist systems, automated intersection management
\cite{fajardo2011automated}, tight-formation platooning, and cooperative
sensing \cite{choi2016mmWaveVehicular, lachapelle2020riskIcassp}, all of which
demand accurate (e.g., 50-cm at 95\%) ground vehicle positioning in an urban
environment.  But the majority of positioning techniques developed thus far
depend on lidar or cameras, which perform poorly in low-visibility conditions
such as snowy whiteout, dense fog, or heavy rain. Adoption of AGVs in many
parts of the world will require all-weather localization techniques.

Radio-wave-based sensing techniques such as radar and GNSS
(global navigation satellite system) remain operable even in extreme weather
conditions~\cite{yen2015evaluation} because their longer-wavelength
electromagnetic radiation penetrates snow, fog, and rain.
Carrier-phase-differential GNSS (CDGNSS) has been successfully applied for the
past two decades as an all-weather decimeter-accurate localization technique in
open-sky conditions. Proprioceptive sensors such as inertial measurement units
(IMUs) also continue to operate regardless of external conditions. Coupling a
CDGNSS receiver with a tactical-grade inertial sensor, as
in~\cite{petovello2004benefits,scherzinger2006precise,
zhangComparisonWithTactical2006,kennedy2006architecture} delivers robust
high-accuracy positioning even during the extended signal outages common in the
urban environment, but such systems are far too expensive for widespread
deployment on AGVs.  Recent work has shown that 20-cm-accurate (95\%) CDGNSS
positioning is possible at low cost even in dense urban areas, but solution
availability remains below 90\%, with occasional long gaps between
high-accuracy solutions~\cite{humphreys2019deepUrbanIts}.  Moreover, the global
trend of increasing radio interference in the GNSS bands, whether accidental or
deliberate~\cite{humphreysGNSShandbook}, underscores the need for
GNSS-independent localization: GNSS jamming cannot be allowed to paralyze an
area's automated vehicle networks.

Clearly, there is a need for AGV localization that is low cost, accurate at the
sub-\num{50}-cm level, robust to low-visibility conditions, and continuously
available.  This paper is the first to establish that low-cost inertial- and
automotive-radar-based localization can meet these criteria.

\begin{figure*}[htb!]
  \centering
  \begin{minipage}[b]{0.245\textwidth}
    \centering
    \includegraphics[width=\linewidth]{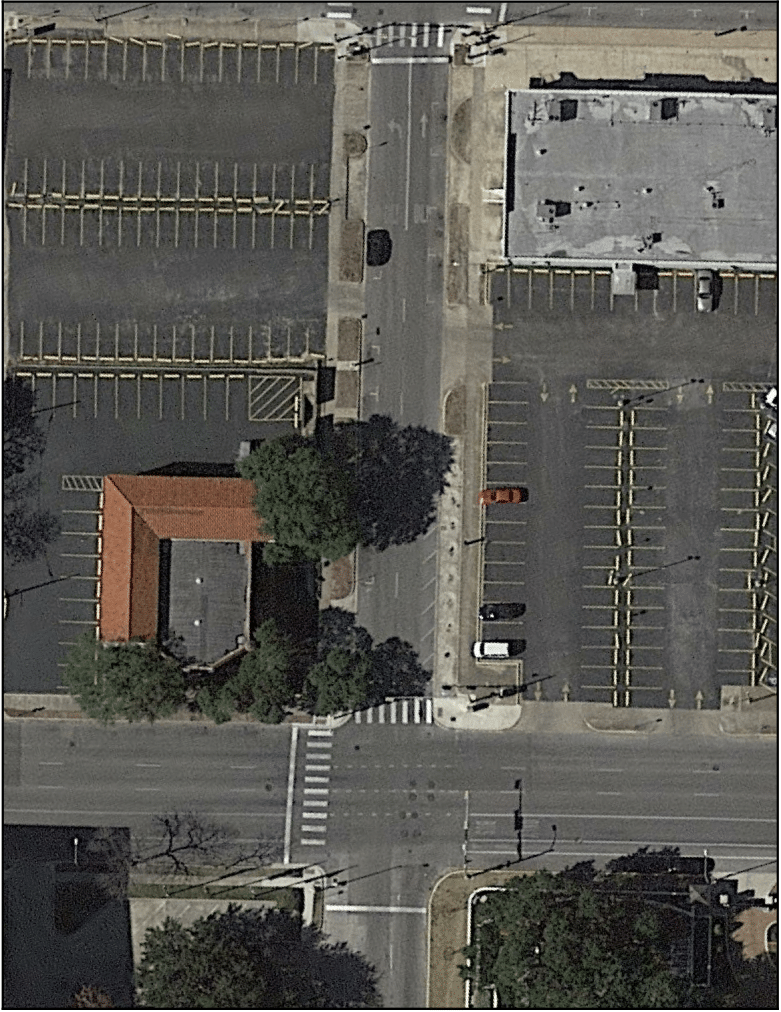}
    \subcaption{}
    \label{fig:satellite-view}
  \end{minipage}
  \begin{minipage}[b]{0.245\textwidth}
    \centering
    \includegraphics[width=\linewidth]{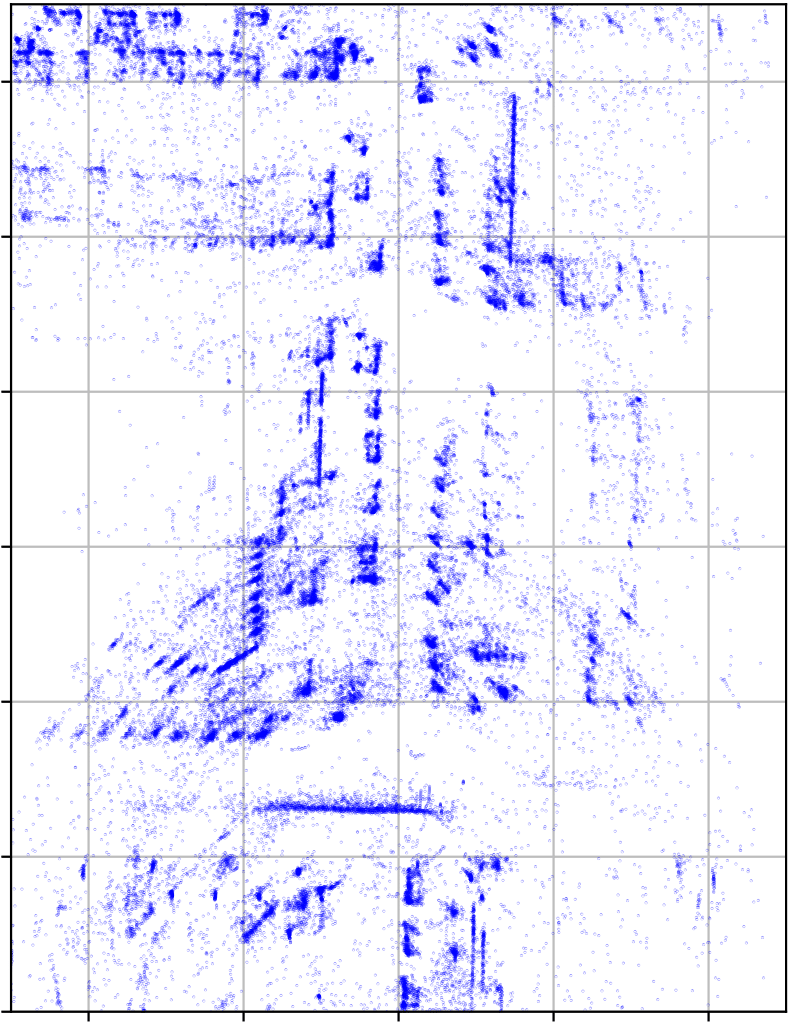}
    \subcaption{}
    \label{fig:radar-map}
  \end{minipage}
  \begin{minipage}[b]{0.245\textwidth}
    \centering
    \includegraphics[width=\linewidth]{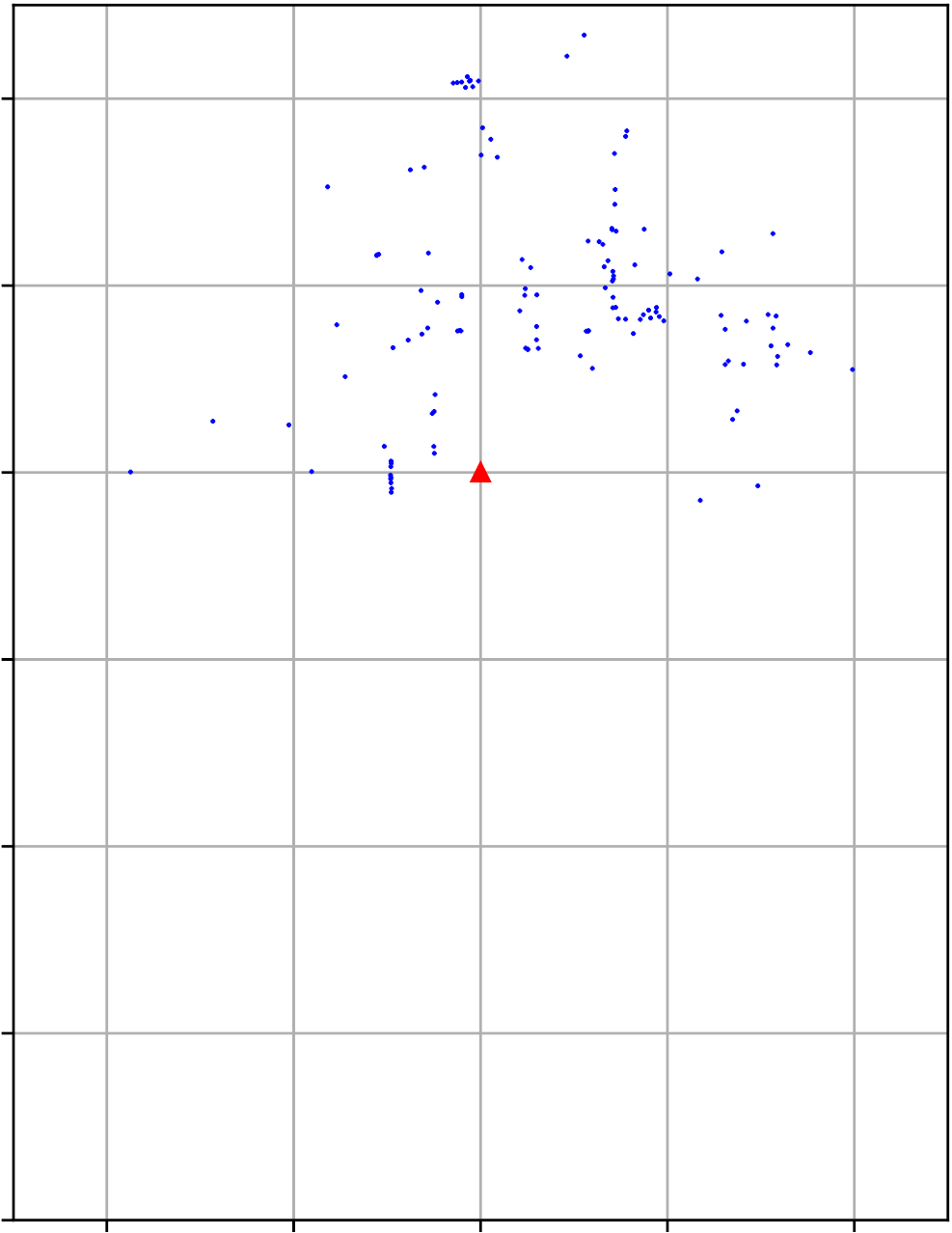}
    \subcaption{}
    \label{fig:single-scan}
  \end{minipage}
  \begin{minipage}[b]{0.245\textwidth}
    \centering
    \includegraphics[width=\linewidth]{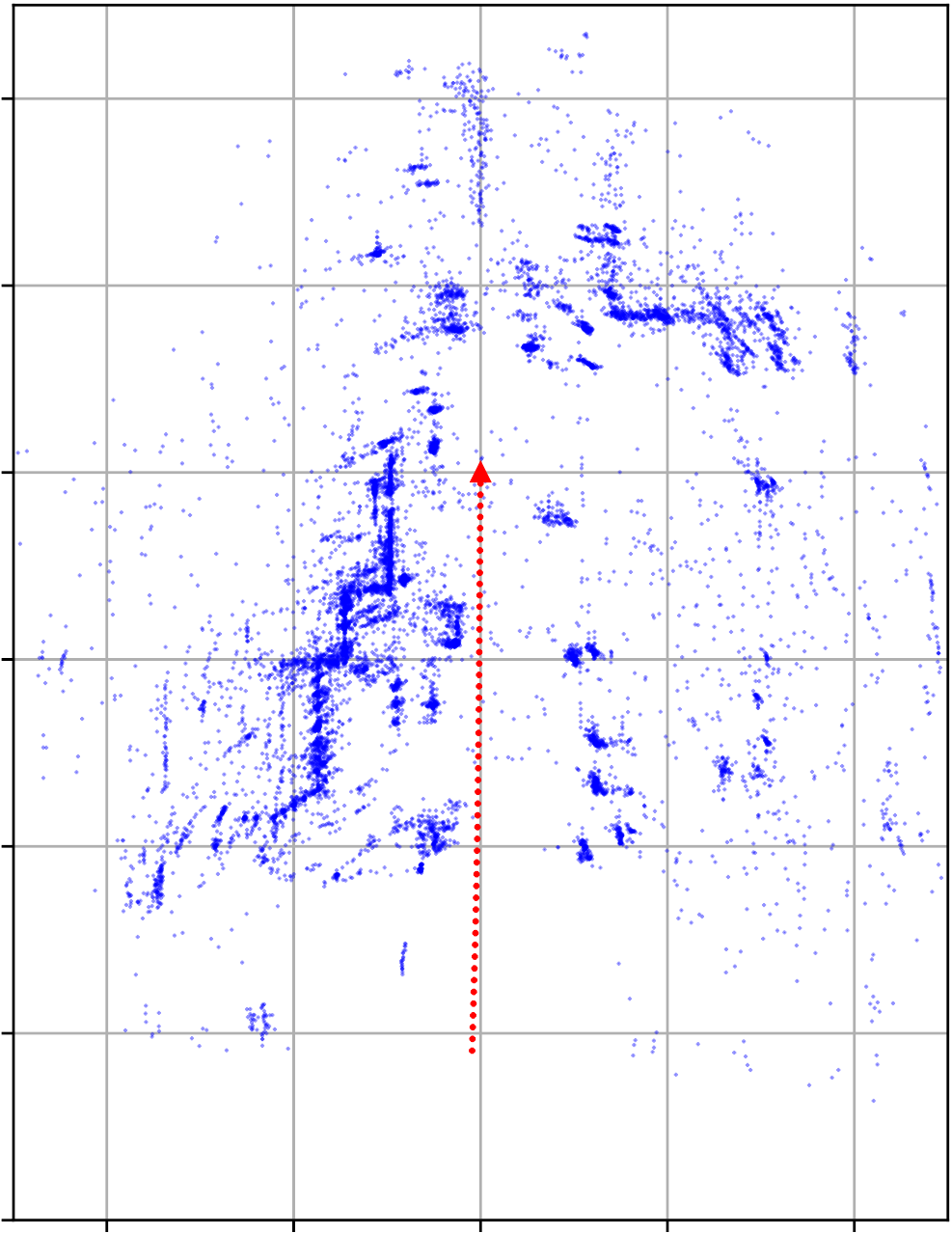}
    \subcaption{}
    \label{fig:radar-batch}
  \end{minipage}
  \caption{Panel (a) shows a satellite view of the environment being mapped
  with automotive radar. Panel (b) shows the generated radar map point cloud
  with vehicle pose obtained from a reference localization system. Note the
  repeating structure along the road side due to parked vehicles. An individual
  radar scan obtained during localization is shown in panel (c), along with the
  red triangle denoting vehicle location and heading. The scan is sparse and
  contains significant clutter, making it challenging to register to the prior
  map. Panel (d) shows a batch of radar scans during localization, with the
  red dots denoting the vehicle trajectory over the past five seconds.  The
  batch captures the underlying structure which can be registered to the
  prior map.}
  \label{fig:radar-scan-matching}
\end{figure*}

Mass-market commercialization has brought the cost of automotive radar down
enough that virtually all current production vehicles are equipped with at
least one radar unit, which serves as the primary sensor for adaptive cruise
control and automatic emergency braking.  But use of automotive radar for
localization faces the significant challenges of data sparsity and noise: an
automotive radar scan has vastly lower resolution than a camera image or a
dense lidar scan, and is subject to high rates of false detection (clutter) and
missed detection.  As such, it is nearly impossible to deduce semantic
information or to extract distinctive environmental features from an individual
radar scan.  This is clear from Fig.~\ref{fig:single-scan}, which shows a
sparse smattering of reflections from a single composite scan using three radar
units. The key to localization is to aggregate sequential scans into a batch,
as in Fig.~\ref{fig:radar-batch}, where environmental structure is clearly
evident.  Even still, the data remain so sparse that localization based on
traditional machine vision feature extraction and matching is not promising.
Additionally, stable short-term odometry is a pre-requisite for aggregating
radar scans, which in itself is a challenge when dealing with low-cost inertial
sensing.

This paper proposes a two-step process for radar-based localization.  The first
is the mapping step: creation of a geo-referenced two-dimensional aggregated
map of all radar targets across an area of interest.  Fig.~\ref{fig:radar-map}
shows such a map, hereafter referred to as a radar map. The full radar map
used throughout this paper, of which Fig.~\ref{fig:radar-map} is a part, was
constructed with the benefit of a highly stable inertial platform so that a
trustworthy ground truth map would be available against which maps generated by
other techniques could be compared.  But an expensive inertial system or
dedicated mobile mapping vehicle is not required to create a radar map.
Instead, it can be crowd-sourced from the very user vehicles that will
ultimately exploit the map for localization.  During periods of favorable
lighting conditions and good visibility, user vehicles can exploit a
combination of low-cost CDGNSS, as in~\cite{humphreys2019deepUrbanIts}, and
GNSS-aided visual simultaneous localization and mapping, as
in~\cite{narula2018accurate}, to achieve the continuous
decimeter-and-sub-degree-accurate geo-referenced position and orientation
(pose) required to lay down an accurate radar map.  In other words, the radar
map can be \emph{created} when visibility is good and then \emph{exploited} at
any later time, such as during times of poor visibility.

Despite aggregation over multiple vehicle passes, the sparse and cluttered
nature of automotive radar data is evident from the radar map shown in
Fig.~\ref{fig:radar-map}: the generated point cloud is much less dense and has
a substantially higher fraction of spurious returns than a typical
lidar-derived point cloud, making automotive-radar-based localization a
significantly more challenging problem.

The second step of this paper's technique is the localization step. Using a
combination of all-weather odometric techniques such as inertial sensing, radar
odometry, and ground vehicle dynamics constraints, a sensor fusion filter
continually tracks the changes in vehicle pose over time. Over the latest short
interval (e.g., \SI{5}{\second}), pose estimates from the filter are used to
spatially organize the multiple radar scans taken over the interval and
generate what is hereafter referred to as a batch of scans, or batch for short.
Fig.~\ref{fig:radar-batch} shows a five-second batch terminating at the same
location as the individual scan in Fig.~\ref{fig:single-scan}. In contrast to
the individual scan, some environmental structure emerges in the batch of
scans, making robust registration to the map feasible. Even so, the
localization problem remains challenging due to the dynamic radar environment:
note the absence of parked cars on the left side of the street during
localization.  The batch of scans is matched against the prior map of the
surroundings to estimate the pose offset of the batch from the truth. This pose
offset is then applied as a measurement to the sensor fusion filter to correct
odometric drift.

\textbf{Contributions.}  This paper's overall contribution is a robust pipeline
for all-weather sub-50-cm urban ground vehicle positioning.  This pipeline
incorporates a computationally-efficient correlation-maximization-based
globally-optimal radar scan registration algorithm that estimates a
two-dimensional translational and a one-dimensional rotational offset between a
prior radar map and a batch of current scans.  Significantly, the registration
algorithm can be applied to the highly sparse and cluttered data produced by
commercially-available low-cost automotive radars.  Maximization of correlation
is shown to be equivalent to minimization of the $L^2$ distance between the
prior map and the batch probability hypothesis densities.  The pipeline
supports the construction of the radar registration estimate and optimally
fuses it with inertial measurements, radar range rate measurements, ground
vehicle dynamics constraints, and cm-accurate GNSS measurements, when
available.  A novel technique for online estimation of the vehicle center of
rotation is introduced, and calibration of various other extrinsic parameters
necessary for optimal sensor fusion is described.

This paper also presents a thorough evaluation of the positioning pipeline
on the large-scale dataset described in~\cite{narula2020texcup}. Data from
automotive sensors are collected over two \SI{1.5}{\hour} driving sessions
through the urban center of Austin, TX on two separate days specifically chosen
to provide variety in traffic and parking patterns.  Comparison with a
post-processed ground truth trajectory shows that proposed pipeline maintains
$95^{\rm th}$-percentile errors below \SI{35}{\centi\meter} in horizontal position
and \ang{0.5} in heading during \SI{60}{\minute} of GNSS-denied driving.

A preliminary version of this paper describing the radar scan registration
algorithm was published in \cite{narula2020radarpositioning}.  The current
version develops and tests a full sensor fusion pipeline that includes the
radar batch estimation as a sub-component.

\textbf{Organization of the rest of this paper.} Sec.~\ref{sec:related-work}
establishes the significance of this contribution in view of the prior work in
related fields. The radar batch-based pose estimation technique for the
low-cost automotive radar sensor model is developed in
Sec.~\ref{sec:batchproc}.  Sec.~\ref{sec:fusion} describes the overall sensor
fusion architecture involving inertial sensing, GNSS, motion constraints, and
radar measurements.  Implementation details and experimental results from field
evaluation are presented in Sec.~\ref{sec:results}, and
Sec.~\ref{sec:conclusion} provides concluding remarks.


\section{Related Work}
\label{sec:related-work}

This section reviews a wide variety of literature on mapping and localization
with radar and radar-inertial sensing. This includes prior work on point cloud
alignment and image registration techniques, occupancy grid-based mapping and
localization, random-finite-set-based mapping and localization, and
inertial-aided mapping and localization.

\textbf{Related work in point cloud alignment.} A radar-based map can have many
different representations.  One obvious representation is to store all the
radar measurements as a point cloud.  Fig.~\ref{fig:radar-map} is an example of
this representation.  Localization within this map can be performed with point
cloud registration techniques like the iterative closest point (ICP) algorithm.
ICP is known to converge to local minima which may occur due to outlying points
that do not have correspondences in the two point clouds being aligned. A
number of variations and generalizations of ICP robust to such outliers have
been proposed in the literature~\cite{chetverikov2002trimmed, ward2016vehicle,
holder2019real, tsin2004correlation, jian2010robust, myronenko2010point,
gao2019filterreg}.  A few of these have been applied specifically to automotive
radar data~\cite{ward2016vehicle, holder2019real}.  But the technique
in~\cite{ward2016vehicle} is only evaluated on a \SI{5}{\minute} dataset,
while~\cite{holder2019real} performs poorly on datasets larger than
\SI{1}{\kilo\meter}.

This paper steers away from ICP and its gradient-based variants because
automotive radar data in urban areas exhibit another source of
incorrect-but-plausible registration solutions which are not addressed in the
above literature---repetitive structure, e.g., due to a series of parked cars,
may result in multiple locally-optimal solutions within
\SIrange[range-phrase=--,range-units=single]{2}{3}{\meter} of the
globally-optimal solution. Gradient-based techniques which iteratively estimate
correspondences based on the distance between pairs of points are susceptible
to converge to such locally-optimal solutions. Accordingly, the batch-based
pose estimator proposed in this paper is designed to approximate the
globally-optimal solution.

In contrast to ICP and its variants, globally-optimal point cloud registration
can be achieved by performing global point correspondence based on distinctive
feature descriptors~\cite{cen2018precise, cen2019radar, barnes2020under}. All of
these works use a sophisticated mechanically-rotating radar unit that is not
expected to be available on an AGV. Feature description and matching on the
low-cost automotive radars used in this paper is likely to be fragile.  Even
when using the mechanically-rotating radar,~\cite{barnes2019masking} shows that
a correlation-based approach, such as the one developed in this paper,
outperforms other feature-descriptor-based point cloud methods.

\textbf{Related work in image registration and occupancy grid techniques.}
Occupancy grid mapping and localization techniques have been traditionally
applied for lidar-based systems, and recent work in~\cite{schuster2016landmark,
schoen2017real} has explored similar techniques with automotive radar data. In
contrast to batch-based pose estimation described in this paper,
both~\cite{schuster2016landmark} and~\cite{schoen2017real} perform
particle-filter based localization with individual scans, as is typical for
lidar-based systems. These methods were only evaluated on small-scale datasets
collected in a parking lot, and even so, the reported meter-level localization
accuracy is not sufficient for lane-level positioning.

Occupancy grid maps are similar to camera-based top-down images, and thus may
be aligned with image registration techniques, that may be
visual-descriptor-based~\cite{callmer2011radar, hong2020radarslam} or
correlation-based~\cite{yoneda2018vehicle}. Reliable extraction and matching of
visual features, e.g., SIFT or SURF, is significantly more challenging with
automotive radar data. Correlation-based registration is more
robust~\cite{yoneda2018vehicle, barnes2019masking}, and forms the basis of one
of the components in this paper. In contrast to prior
work~\cite{yoneda2018vehicle, barnes2019masking}, this paper provides a
probabilistic interpretation for the correlation operation. The
mechanically-rotating radar of~\cite{barnes2019masking} allows
correlation-based pose estimation based on a single scan of radar data. But for
the low-cost automotive radars used in this paper, it becomes necessary to
accumulate radar scans over time, which requires integration with other
odometric sensors. This paper develops and demonstrates a complete sensor
fusion pipeline around radar-based pose estimation and evaluates its
performance on a large urban dataset.

\textbf{Related work in random finite set mapping and localization.} The
occupancy grid representation commonly used in robotics is an approximation to
the probability hypothesis density (PHD)
function~\cite{mahler2003multitargetFirstOrder, erdinc2009bin}: a concept first
introduced in the random finite set (RFS) based target tracking literature.
Unsurprisingly, PHD- and RFS-based mapping and localization have been
previously studied in~\cite{mullane2011random, deusch2015labeled,
stubler2017continuously}. In contrast to occupancy grid-based methods,
techniques in~\cite{mullane2011random, deusch2015labeled,
stubler2017continuously} make the point target assumption where no target may
generate more than one measurement in a single scan, and no target may occlude
another target.  However, in reality, planar and extended targets such as walls
and building fronts are commonplace in the urban AGV environment. Mapping of
ellipsoidal extended targets has recently been proposed
in~\cite{fatemi2017poisson}, but occlusions are not modeled and only simulation
results are presented.

\textbf{Related work in inertial-aided mapping and localization.} This paper
couples radar batch-based pose estimation with other all-weather automotive
sensors such as IMU and GNSS. Inertial aiding has been widely applied in
visual- and lidar-based mapping and localization~\cite{qin2018vins,
mur2017visual, chiang2020navigation, forster2013collaborative,
steder2008visual, ye2019tightly, li2014lidar}. This paper extends
inertial-aiding to sensors that can operate under harsh weather conditions.
Recently, radar measurements have been applied to constrain IMU position drift
in~\cite{barra2019localization}. Radar-inertial odometry for indoor robots has
been proposed in~\cite{almalioglu2019milli, kramer2020radar}. This paper is the
first to integrate low-cost automotive radars with inertial sensing, GNSS, and
ground vehicle dynamics for lane-level accurate positioning in challenging
urban environments.

\section{Radar-Batch-Based Pose Estimation}
\label{sec:batchproc}

This section describes the formulation of the radar-batch-based pose estimation
method introduced in this paper. It first details the statistical motivation
behind the method, and then develops an efficient approximation to the
globally-optimal estimator. The output of this estimator acts as one of the
measurements provided to the overall localization system presented later in
Sec.~\ref{sec:fusion}.

\subsection{Pose Estimation using Probability Hypothesis Density}
\label{sec:phd-localization}

For the purpose of radar-based pose estimation, an AGVs environment can be
described as a collection of arbitrarily shaped radar reflectors in a specific
spatial arrangement. Assuming sufficient temporal permanence of this
environment, radar-equipped AGVs make sample measurements of the underlying
structure over time.

\subsubsection{The Probability Hypothesis Density Function}

A probabilistic description of the radar environment is required to set up
radar-based pose estimation as an optimization problem. This paper chooses the
PHD function~\cite{mahler2003multitargetFirstOrder} representation of the radar
environment. The PHD at a given location gives the density of the expected
number of radar reflectors at that location. For a static radar environment,
the PHD $D(\bm{x})$ at a location $\bm{x} \in \mathcal{X}$ can be written as
\[
  D(\bm{x}) = I \cdot p(\bm{x})
\]
where $\mathcal{X}$ is the set of all locations in the environment, $p(\bm{x})$
is a probability density function such that $\int p(\bm{x}) {\rm d}\bm{x} = 1$,
and $I$, the intensity, is the total number of radar reflectors in the
environment. For a time-varying radar environment, both $I$ and $p(\bm{x})$ are
functions of time. For $\mathcal{A} \subset \mathcal{X}$, the expected number
of radar reflectors in $\mathcal{A}$ is given as
\[
  I_\mathcal{A} = \int_\mathcal{A} D(\bm{x}) {\rm d}\bm{x}
\]

\subsubsection{Estimating Vehicle State from PHDs}

Let $D_{\rm m}(\bm{x})$ denote the ``map'' PHD function representing the
distribution of radar reflectors in an environment, estimated as a result of
mapping with known vehicle poses. During localization, the vehicle makes a
radar scan, or a series of consecutive radar scans. A natural solution to the
pose estimation problem may be stated as the vehicle pose which maximizes
the likelihood of the observed batch of scans, given that the scan was drawn
from $D_{\rm m}(\bm{x})$~\cite{myronenko2010point}. This maximum likelihood
estimate (MLE) has many desirable properties such as asymptotic efficiency.
However, the MLE solution is known to be sensitive to outliers that may occur
if the batch of scans was sampled from a slightly different PHD, e.g., due to
variations in the radar environment between mapping and
localization~\cite{jian2010robust}.

A more robust solution to the PHD-based pose estimation problem may be stated
as follows.  Let $\bm{\Theta}$ denote the vector of parameters of the rigid or
non-rigid transformation $\mathcal{T}$ between the vehicle's prior belief of
its pose, and its true pose. For example, in case of a two-dimensional rigid
transformation, $\bm{\Theta} = {\left[ \Delta x, \Delta y, \Delta \phi
\right]}^\top$, where $\Delta x$ and $\Delta y$ denote a two-dimensional
position and $\Delta \phi$ denotes heading.  Also, let $D_{\rm
b}(\bm{x}^\prime)$ denote a local ``batch'' PHD function estimated from a batch
of scans during localization, defined over $\bm{x}^\prime \in \mathcal{A}
\subset \mathcal{X}$.  This PHD is represented in the coordinate system
consistent with vehicle's prior belief, such that $\bm{x}^\prime =
\mathcal{T}_{\bm{\Theta}}(\bm{x})$.  Estimating the vehicle pose during
localization is defined as estimating $\bm{\Theta}$ such that some distance
metric between the PHDs $D_{\rm m}(\bm{x})$ and $D_{\rm b}(\bm{x}^\prime)$ is
minimized.

This paper chooses the $L^2$ distance between $D_{\rm m}(\bm{x})$ and $D_{\rm
f}(\bm{x}^\prime)$ as the distance metric to be minimized. As compared to the
MLE which minimizes Kullback-Leibler divergence, $L^2$ minimization trades off
asymptotic efficiency for robustness to measurement model
inaccuracy~\cite{jian2010robust}. The $L^2$ distance $d_{L^2}(\bm{\Theta})$ to
be minimized is given as
\[
  d_{L^2}(\bm{\Theta}) = \int_\mathcal{A} {\left( D_{\rm m}(\bm{x}) - D_{\rm b}(\mathcal{T}_{\bm{\Theta}}(\bm{x})) \right)}^2 {\rm d}\bm{x}
\]

For rigid two-dimensional transformations, it can be shown as follows that
minimizing the $L^2$ distance between the PHDs is equivalent to maximization of
the cross-correlation between the PHDs.
\begin{align*}
  \widehat{\bm{\Theta}} &= \argmin_{\bm{\Theta}^\prime} \int_\mathcal{A} {\left( D_{\rm m}(\bm{x}) - D_{\rm b}(\mathcal{T}_{\bm{\Theta}^\prime}(\bm{x})) \right)}^2 {\rm d}\bm{x} \\
  &
  \begin{multlined}[b]
    = \argmin_{\bm{\Theta}^\prime} \left[ \int_\mathcal{A} D_{\rm m}^2(\bm{x}) {\rm d}\bm{x} + \int_\mathcal{A} D_{\rm b}^2(\mathcal{T}_{\bm{\Theta}^\prime}(\bm{x})) {\rm d}\bm{x} \right. \\
                                   \left. -2\int_\mathcal{A} D_{\rm m}(\bm{x}) D_{\rm b}(\mathcal{T}_{\bm{\Theta}^\prime}(\bm{x})) {\rm d}\bm{x} \right]
  \end{multlined}
\end{align*}
Note that the first term above is fixed during optimization, while the second
term is invariant under rigid transformation.  As a result, the above
optimization is equivalent to maximizing the cross-correlation:
\begin{equation}
  \widehat{\bm{\Theta}} = \argmax_{\bm{\Theta}^\prime} \int_\mathcal{A} D_{\rm m}(\bm{x}) D_{\rm b}(\mathcal{T}_{\bm{\Theta}^\prime}(\bm{x})) {\rm d}\bm{x}
  \label{eq:phd-correlation}
\end{equation}
For differentiable $D_{\rm m}$ and $D_{\rm b}$, the above optimization can be
solved with gradient-based methods. However, the cross-correlation maximization
problem in the urban AGV environment may have locally optimal solutions in the
vicinity of the global minimum due to repetitive structure of radar reflectors.
In applications with high integrity requirements, a search for the globally
optimal solution is necessary. This paper notes that if the PHDs in
(\ref{eq:phd-correlation}) were to be discretized in $\bm{x}$, then the
cross-correlation values can be evaluated exhaustively with computationally
efficient techniques. Let $\bm{x}_{pq}$ denote the location at the $(p,q)$
translational offset in discretized $\mathcal{A}$. Then
\begin{equation}
  \widehat{\bm{\Theta}} = \argmax_{\bm{\Theta}^\prime} \sum_{p=0}^{P-1}
  \sum_{q=0}^{Q-1} D_{\rm m}(\bm{x}_{pq}) D_{\rm b}(\round{\mathcal{T}_{\bm{\Theta}^\prime}(\bm{x}_{pq})})
  \label{eq:phd-discrete-correlation}
\end{equation}
where $\round{.}$ denotes the nearest grid point in the discretized space.

The technique developed above relies on the PHDs $D_{\rm m}$ and $D_{\rm b}$.
The next subsections detail the recipe for estimating these PHDs from the radar
observations.

\subsection{Estimating the map PHD from measurements}
\label{sec:phd-approximation}

This section addresses the procedure to estimate the map PHD $D_{\rm
m}(\bm{x})$ from radar measurements. This paper works with an occupancy grid
map (OGM) approximation to the continuous PHD function.
In~\cite{erdinc2009bin}, it has been shown that the PHD representation is a
limiting case of the OGM as the grid cell size becomes vanishingly small.
Intuitively, let $c_{pq}$ denote the grid cell region with center
$\bm{x}_{pq}$, and let $\delta c_{pq}$ denote the area of this grid cell, which
is small enough such that no more than one reflector may be found in any cell.
Let $p_{pq} (O)$ denote the occupancy probability of $c_{pq}$, and let
$\mathcal{A}$ be defined as the region formed by the union of all $c_{pq}$
whose centers $\bm{x}_{pq}$ fall within $\mathcal{A}$. Then, the expected
number of radar reflectors $\mathbb{E} \left[ \abs{\mathcal{A}} \right]$ in
$\mathcal{A}$ is given by
\begin{align*}
  \mathbb{E} \left[ | \mathcal{A} | \right] = \sum_{c_{pq} \in \mathcal{A}} p_{pq}(O)
  &= \sum_{c_{pq} \in \mathcal{A}} \frac{p_{pq}(O)}{\delta c_{pq}} \delta c_{pq} \\
  &\triangleq \sum_{c_{pq} \in \mathcal{A}} \bar{D}(\bm{x}_{pq}) \delta c_{pq} \\
  &= \int_\mathcal{A} \bar{D}(\bm{x}_{pq}) {\rm d}\bm{x}, \quad {\rm as}~\lim_{\delta c_{pq} \rightarrow 0}
\end{align*}
where $\bar{D}(\bm{x}_{pq}) \equiv \frac{p_{pq}(O)}{\delta c_{pq}}$ can be
considered to be an approximation of the PHD $D(\bm{x})$ for $\bm{x} \in
c_{pq}$ since its integral over $\mathcal{A}$ is equal to the expected number
of reflectors in $\mathcal{A}$.

The advantage of working with an OGM approximation of the PHD is two-fold:
first, since the OGM does not attempt to model individual objects, it is
straightforward to represent arbitrarily-shaped objects, and second, in
contrast to the ``point target'' measurement model assumption in standard PHD
filtering, the OGM can straightforwardly model occlusions due to extended
objects.

At this point, the task of estimating $D_{\rm m}(\bm{x})$ has been reduced to
estimating the occupancy probability of each grid cell in discretized
$\mathcal{A}$. Each grid cell $c_{pq}$ takes up one of two states: occupied
($O$) or free ($F$). Based on the radar measurement $\bm{z}_k$ at each time
$k$, the Bernoulli probability distribution of such binary state cells may be
recursively updated with the binary Bayes filter. In particular, let
$\bm{z}_{1:k}$ denote all radar measurements made up to time $k$, and let
\begin{equation}
  l_{pq}^k(O) \equiv \log{\frac{p_{pq}(O~|~\bm{z}_{1:k})}{1 - p_{pq}(O~|~\bm{z}_{1:k})}}
  \label{eq:log-odds}
\end{equation}
denote the \emph{log odds ratio} of $c_{pq}$ being in state $O$. Also define
$l_{pq}^0(O)$ as
\[
  l_{pq}^0(O) \equiv \log{\frac{p_{pq}(O)}{1 - p_{pq}(O)}}
\]
with $p_{pq}(O)$ being the prior belief on the occupancy state of $c_{pq}$
before any measurements are made. With these definitions, the binary Bayes
filter update is given by~\cite{thrun2005probabilistic}
\begin{equation}
  l_{pq}^k(O) = \log{\frac{p_{pq}(O~|~\bm{z}_k)}{1 - p_{pq}(O~|~\bm{z}_k)}} - l_{pq}^0(O) + l_{pq}^{k-1}(O)
  \label{eq:binary-bayes}
\end{equation}
where $p_{pq}(O~|~\bm{z}_k)$ is known as the \emph{inverse} sensor model:
it describes the probability of $c_{pq}$ being in state $O$, given only the
latest radar scan $\bm{z}_k$.

The required occupancy probability $p_{pq}(O~|~\bm{z}_{1:k})$ is easy to
compute from the log odds ratio in (\ref{eq:log-odds}). Observe that the
inverse sensor model $p_{pq}(O~|~\bm{z}_k)$, in addition to the prior occupancy
belief $p_{pq}(O)$, completely describes the procedure for estimating the OGM
from radar measurements, and hence approximating the PHD. Adapting
$p_{pq}(O~|~\bm{z}_k)$ to the characteristics of the automotive radar sensors,
however, is not straightforward, and is discussed next.

\subsection{Automotive Radar Inverse Sensor Model}

\begin{figure}[ht]
  \centering
  \includegraphics[width=\linewidth] {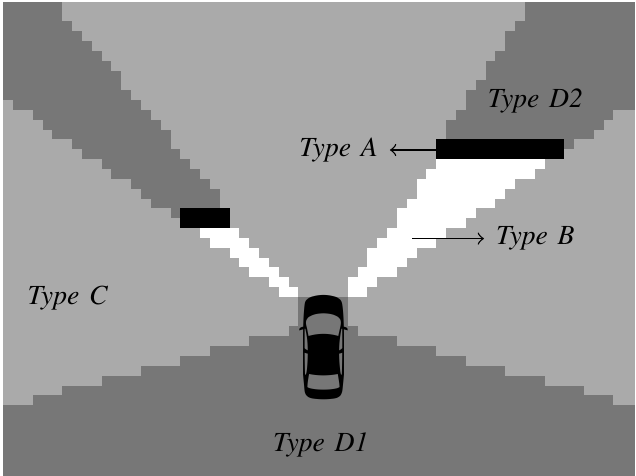}
  \caption{Schematic diagram showing four types of grid cells.}
  \label{fig:inverse-sensor}
\end{figure}

This section addresses the challenge of adapting the inverse sensor model
$p_{pq}(O~|~\bm{z}_k)$ to the measurement characteristics of automotive radar
sensors. Fig.~\ref{fig:inverse-sensor} shows a simplified radar scan $\bm{z}_k$
of an underlying occupancy grid. For clarity of exposition, four distinct
categories of grid cells in Fig.~\ref{fig:inverse-sensor} are defined below:
\begin{itemize}
  \item \emph{Type A}: Grid cells in the vicinity of a radar range-azimuth
    return.
  \item \emph{Type B}: Grid cells along the path between the radar sensor and
    \emph{Type A} grid cells.
  \item \emph{Type C}: Grid cells in the ``viewshed'' of the radar sensor,
    i.e., within the radar field-of-view and not shadowed by another object,
    but not of \emph{Type A} or \emph{Type B}.
  \item \emph{Type D}: Grid cells outside the field-of-view of the radar
    (\emph{Type D1}) or shadowed by other objects closer to the radar
    (\emph{Type D2}).
\end{itemize}
The inverse sensor model must choose a $p_{pq}(O~|~\bm{z}_k)$ value for each of
these types of grid cells. In the following, the subscript $pq$ is dropped for
cleaner notation.

\subsubsection{Conventional Choices for the Inverse Sensor Model}

Since $\bm{z}_k$ provides no additional information on \emph{Type D} grid
cells, the occupancy in these cells is conditionally independent of $\bm{z}_k$,
that is
\[
  p^D(O~|~\bm{z}_k) = p(O)
\]
where $p(O)$ is the prior probability of occupancy defined earlier in
Sec.~\ref{sec:phd-localization}.

Grid cells of \emph{Type B} and \emph{Type C} may be hypothesized to have low
occupancy probability, since these grid cells were scanned by the sensor but no
return was obtained. As a result, conventionally
\[
  p^B(O~|~\bm{z}_k) \leq p(O)
\]
and
\[
  p^C(O~|~\bm{z}_k) \leq p(O)
\]

Finally, grid cells of \emph{Type A} may be hypothesized to have higher
occupancy probability, since a return has been observed in the vicinity of
these cells. Conventionally,
\[
  p^A(O~|~\bm{z}_k) \geq p(O)
\]
In the limit, if
the OGM grid cell size is comparable to the sensor range and angle uncertainty,
or if the number of scans is large enough such that the uncertainty is captured
empirically, only the grid cells that contain the sensor measurement may be
considered to be of \emph{Type A}.

\subsubsection{Automotive Radar Sensor Characteristics}

Intense clutter properties and sparsity of the automotive radar data complicate
the choice of the inverse sensor model.

\textbf{Sparsity.} First, sparsity of the radar scan implies that many occupied
\emph{Type A} grid cells in the radar environment might be incorrectly
categorized as free \emph{Type C} cells.  This can be observed in
Fig.~\ref{fig:radar-scan-matching}. As evidenced by the batch of scans in
Fig.~\ref{fig:radar-batch}, the radar environment is ``dense'' in that many
grid cells contain radar reflectors. However, any individual radar scan, such
as the one shown in Fig.~\ref{fig:single-scan}, suggests a much more sparse
radar environment. As a result, a grid cell which is occupied in truth will be
incorrectly categorized as \emph{Type C} in many scans, and correctly as
\emph{Type A} in a few scans.

The sparsity of radar returns also makes it challenging to distinguish
\emph{Type C} cells from cells of \emph{Type D2}. Since many occluding
obstacles are not detected in each scan, the occluded cells of \emph{Type D2}
are conflated with free cells of \emph{Type C}.

In context of the inverse sensor model, as the radar scan becomes more sparse
\[
  p^C(O~|~\bm{z}_k) \rightarrow {p^D(O~|~\bm{z}_k)}^{-}
\]
where the superscript $-$ denotes a limit approaching from below. Intuitively,
approaching $p^D(O~|~\bm{z}_k)$ implies that the measurement $\bm{z}_k$ is very
sparse in comparison to the true occupancy, and thus does not provide much
information on lack of occupancy.

\textbf{Clutter.} Second, there is the matter of clutter. The grid cells in the
vicinity of a clutter measurement may be incorrectly categorized as \emph{Type
A}, and the grid cells along the path between the radar and clutter measurement
may be incorrectly categorized as \emph{Type B}.

In context of the inverse sensor model, as the radar scan becomes more
cluttered
\begin{align*}
  p^B(O~|~\bm{z}_k) &\rightarrow {p^D(O~|~\bm{z}_k)}^{-} \\
  p^A(O~|~\bm{z}_k) &\rightarrow {p^D(O~|~\bm{z}_k)}^{+}
\end{align*}
where the superscript $+$ denotes a limit approaching from above.

\subsubsection{A Pessimistic Inverse Sensor Model}

The results presented in Sec.~\ref{sec:results} are based on a pessimistic
sensor model, such that $p^B(O~|~\bm{z}_k) = p^C(O~|~\bm{z}_k) =
p^D(O~|~\bm{z}_k)$. This model assumes that the radar measurements provide no
information about free space in the radar environment.

In particular, the inverse sensor model assumes
\[
  p^B(O~|~\bm{z}_k) = p^C(O~|~\bm{z}_k) = p^D(O~|~\bm{z}_k) = p(O) = 0.1
\]
and
\[
  p^A(O~|~\bm{z}_k) = 0.2
\]

\subsection{Estimating the batch PHD from measurements}

The procedure for generating an approximation to $D_{\rm b}(\bm{x}^\prime)$
from a batch of radar measurements is identical to the procedure for generating
$D_{\rm m}(\bm{x})$ from mapping vehicle data, except that precise, absolute
location and orientation data is not available during localization.  Instead,
pose estimates from the sensor fusion filter described in Sec.~\ref{sec:fusion}
are used to estimate the relative locations and orientations of each radar scan
in the batch, and the scans are transformed into a common coordinate frame
before updating the occupancy state of grid cells.

Once the map and batch PHDs have been approximated from radar measurements, the
correlation-maximization technique developed in Sec.~\ref{sec:phd-localization}
can be applied to obtain the estimate $\widehat{\bm{\Theta}}$. This estimate is
handed back to the sensor fusion filter as a pose offset measurement to
constrain the odometric drift during absence of other sources of absolute
localization, e.g., GNSS.


\section{State Estimation with Sensor Fusion}
\label{sec:fusion}

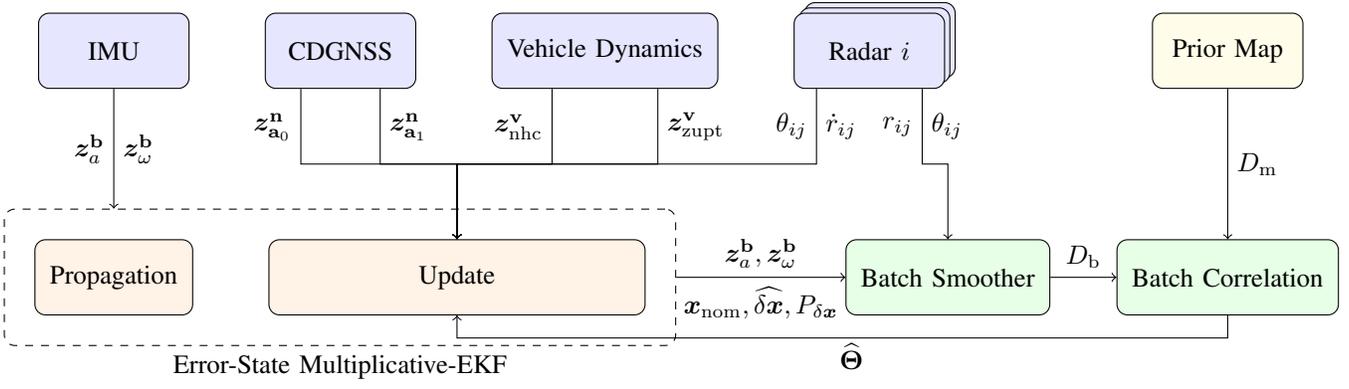
\begin{figure*}[ht]
  \centering
  \tikzstyle{measurement} = [rectangle, rounded corners, minimum width=2cm, minimum height=1cm, inner sep=0.2cm, text centered, draw=black, fill=blue!10]
\tikzstyle{filter} = [rectangle, rounded corners, minimum width=2cm, minimum height=1cm, inner sep=0.2cm, text centered, draw=black, fill=orange!10]
\tikzstyle{batch} = [rectangle, rounded corners, minimum width=2cm, minimum height=1cm, inner sep=0.2cm, text centered, draw=black, fill=green!10]
\tikzstyle{map} = [rectangle, rounded corners, minimum width=2cm, minimum height=1cm, inner sep=0.2cm, text centered, draw=black, fill=yellow!10]

\begin{tikzpicture}

  \node (imu) [measurement] {IMU};

  \node (cdgnss) [measurement, right=of imu] {CDGNSS};

  \node (dynamics) [measurement, right=of cdgnss] {Vehicle Dynamics};

  \node (radars) [measurement, right=of dynamics,
                  copy shadow = {shadow xshift = 4pt, shadow yshift = 4pt},
                  copy shadow = {shadow xshift = 2pt, shadow yshift = 2pt}] {Radar $i$};

  \node (map) [map, right=2.75cm of radars] {Prior Map};

  \node (prop) [filter, below=2cm of imu] {Propagation};
  \node (updt) [filter, minimum width=5cm, right=of prop] {Update};
  \node (mekf) [fit=(prop)(updt), draw, rounded corners, dashed, inner sep=0.4cm, label=below:{Error-State Multiplicative-EKF}] {};

  \node (rtss) [batch, right=2.25cm of mekf] {Batch Smoother};

  \node (corr) [batch, below=2cm of map] {Batch Correlation};

  \draw[->] (imu.south) -- node [left] {$\bm{z}_{a}^{\brm{b}}$} node [right] {$\bm{z}_{\omega}^{\brm{b}}$} (mekf.north -| prop.north);

  \draw[->] ([xshift=-15pt]cdgnss.south) -- node [left] {$\bm{z}_{\brm{a}_0}^{\brm{n}}$} ++(0,-1.0) -| (updt.north);
  \draw[->] ([xshift=+15pt]cdgnss.south) -- node [right] {$\bm{z}_{\brm{a}_1}^{\brm{n}}$} ++(0,-1.0) -| (updt.north);

  \draw[->] ([xshift=-20pt]dynamics.south) -- node [left] {$\bm{z}^{\brm{v}}_{\mathrm{nhc}}$} ++(0,-1.0) -| (updt.north);
  \draw[->] ([xshift=+20pt]dynamics.south) -- node [right] {$\bm{z}^{\brm{v}}_{\mathrm{zupt}}$} ++(0,-1.0) -| (updt.north);

  \draw[->] ([xshift=-20pt]radars.south) -- node [right] {$\dot{r}_{ij}$} node [left] {$\theta_{ij}$} ++(0,-1.0) -| (updt.north);

  \draw[->] ([xshift=+20pt]radars.south) -- node [right] {$\theta_{ij}$} node [left] {$r_{ij}$} ++(0,-1.0) -| (rtss.north);

  \draw[->] (mekf.east) -- node [above] {$\bm{z}_{a}^{\brm{b}}, \bm{z}_{\omega}^{\brm{b}}$} node [below] {$\bm{x}_{\mathrm{nom}}, \widehat{\delta\bm{x}}, P_{\delta\bm{x}}$} (rtss.west);

  \draw[->] (map.south) -- node [right] {$D_{\mathrm{m}}$} (corr.north);
  \draw[->] (rtss.east) -- node [above] {$D_{\mathrm{b}}$} (corr.west);

  \draw[->] (corr.south) -- node [below, xshift=-5cm, yshift=-0.1cm] {$\widehat{\bm{\Theta}}$} ++(0,-0.3) -| (updt.south);

\end{tikzpicture}
  \caption{Block diagram of the localization pipeline. A low-cost MEMS IMU
    provides high-rate specific force and angular rate measurements.  The
    error-state multiplicative extended Kalman filter (EKF) makes use of
    cm-accurate CDGNSS position measurements whenever such measurements are
    available, e.g., in clear-sky GNSS environments.  Radial velocity and
    bearing measurements from low-cost automotive radars are combined with
    nearly-zero sideslip and vertical speed constraints of a ground vehicle to
    continually track and limit the errors in inertial navigation. Smoothed
    batches of radar scans are correlated with a prior map to limit odometric
    position drift during CDGNSS outages.}
  \label{fig:fusion-arch}
\end{figure*}

Thus far, Sec.~\ref{sec:batchproc} has developed the theory and implementation
of the radar batch correlation measurement, which provides an estimate
$\widehat{\bm{\Theta}}$ of the 3-DoF (degrees-of-freedom) pose offset relative
to the prior map. This section details a localization pipeline that
incorporates the batch measurement update along with an array of other
automotive all-weather sensing modalities to track the full 6-DoF vehicle pose
trajectory.  The high-rate pose estimates from this pipeline are also used to
spatially organize individual scans to form the batch of radar scans used in
the batch correlation update.

The choice of sensors available for all-weather localization is limited to
radio-frequency sensors such as GNSS and automotive radars, and to
proprioceptive sensors such as IMUs and wheel encoders. Any additional domain
knowledge, such as properties of ground vehicle dynamics, may also be combined
with these sensor measurements.

The localization pipeline in this paper is developed around a low-cost MEMS
IMU. Fig.~\ref{fig:fusion-arch} shows a block diagram of the overall pipeline.
The error-state multiplicative extended Kalman filter (EKF) makes use of
cm-accurate CDGNSS position measurements whenever such measurements are
available, e.g., in clear-sky GNSS environments.  Radial velocity and bearing
measurements from low-cost automotive radars are combined with nearly-zero
sideslip and vertical speed constraints of a ground vehicle to continually
track and limit the errors in inertial navigation. Smoothed batches of radar
scans are correlated with a prior map to limit odometric position drift during
CDGNSS outages. The following subsections outline the formulation of the
estimator, the nonlinear state dynamics, the various measurement models, and
the necessary calibration procedures.

\subsection{Sensor Platform \& Coordinate Frames}

\begin{figure}[ht]
  \centering
  \input{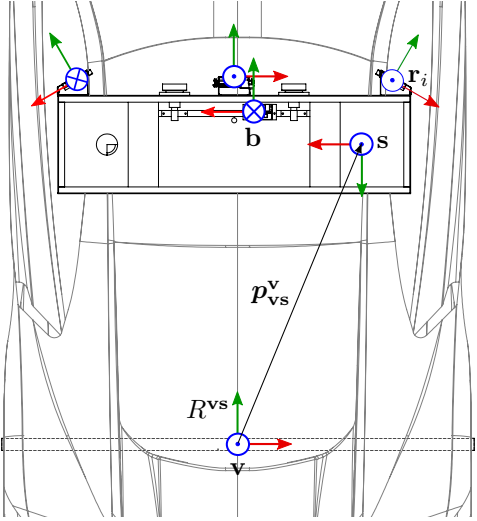}
  \caption{The University of Texas Sensorium is an integrated platform for
  automated and connected vehicle perception research.  It includes three
  automotive radar units, one electronically-scanning radar (ESR) and two
  short-range radars (SRR2s); stereo visible light cameras; automotive- and
  industrial-grade inertial measurement units (IMUs); a dual-antenna,
  multi-frequency software-defined GNSS receiver; and an internal
  computer. An iXblue ATLANS-C CDGNSS-disciplined inertial navigation system
  (INS) (not shown) is mounted at the rear of the platform to provide the
  ground truth trajectory. The vehicle frame $\brm{v}$ is located approximately
  at the center of the line connecting the rear axles.}
  \label{fig:sensorium}
\end{figure}

To facilitate the discussion on measurement models and calibration, the
sensor-instrumented vehicle and a few related coordinate frames are introduced
here. An integrated perception platform called the \emph{Sensorium}, shown
schematically in Fig.~\ref{fig:sensorium}, brings together the various low-cost
automotive sensors considered in this paper. Many of these sensors provide
measurements in their respective local frames, leading to a number of different
coordinate frames that must be considered.

The IMU \emph{body frame}, denoted $\brm{b}$, is the frame defined by the IMU's
accelerometer triad.

The \emph{navigation frame}, denoted $\brm{n}$, is a local geographical
reference frame, e.g., an ENU frame. The estimator wishes to track the pose
trajectory of $\brm{b}$ with respect to $\brm{n}$.

The \emph{radar frames}, denoted $\brm{r}_i$ for the $i$th radar, are local
frames in which the radar sensors report range, range rate, and bearing to a
number of targets.

The \emph{vehicle frame}, denoted $\brm{v}$, is characterized by the direction
in which the vehicle travels when the commanded steering angle is zero. This
direction defines the $y$-axis of $\brm{v}$, as shown in
Fig.~\ref{fig:sensorium}. The origin of $\brm{v}$ is located at the center of
rotation of the vehicle.

The \emph{Sensorium frame}, denoted $\brm{s}$, is defined by the physical
structure of the Sensorium. It is essentially a convenience reference frame in
which the nominal lever arm and orientation between different sensors are
available per the mechanical specifications of the Sensorium. The origin of
$\brm{s}$ is arbitrarily chosen to be co-located with one of the GNSS antennas.

\subsection{Error-State Filtering}
The localization system of Fig.~\ref{fig:fusion-arch} estimates the following
\num{16}-element state vector:
\begin{equation*}
  \bm{x}_k = \left[ \bm{p}_k^{\brm{n}}, \bm{v}_k^{\brm{n}}, \bm{q}_k^{\brm{nb}}, \bm{b}_{a,k}^{\brm{b}}, \bm{b}_{\omega,k}^{\brm{b}} \right]
\end{equation*}
where $\bm{p}_k^{\brm{n}}$ is the vector from $\brm{n}$ to $\brm{b}$ at time
$k$ expressed in $\brm{n}$, $\bm{v}_k^{\brm{n}}$ is the velocity of $\brm{b}$
relative to $\brm{n}$ at time $k$ expressed in the $\brm{n}$ frame,
$\bm{q}_k^{\brm{nb}}$ is the quaternion that rotates a vector from $\brm{b}$ to
$\brm{n}$ at time $k$, and $\bm{b}_{a,k}^{\brm{b}}$ and
$\bm{b}_{\omega,k}^{\brm{b}}$ are the accelerometer and gyroscope biases of the
IMU at time $k$, expressed in $\brm{b}$.

Note that the vehicle orientation only has three effective degrees-of-freedom
since $\bm{q}_k^{\brm{nb}}$ is constrained to be a unit quaternion. Enforcing
such a constraint may result in a singular covariance matrix. This issue is
typically dealt with an error-state filter~\cite{sola2017quaternion} where the
true state is split into a nominal-state vector
\begin{equation*}
  \bm{x}_{\mathrm{nom},k} = \left[ \tilde{\bm{p}}_k^{\brm{n}}, \tilde{\bm{v}}_k^{\brm{n}}, \tilde{\bm{q}}_k^{\brm{nb}}, \tilde{\bm{b}}_{a,k}^{\brm{b}}, \tilde{\bm{b}}_{\omega,k}^{\brm{b}} \right]
\end{equation*}
and an error-state vector $\delta\bm{x}_k$, related by the generalized addition
operator $\oplus$ as follows:
\begin{equation*}
  \bm{x}_k = \bm{x}_{\mathrm{nom},k} \oplus \delta\bm{x}_k
\end{equation*}
where the error-state vector $\delta\bm{x}_k$ is the minimal \num{15}-element
state representation denoted component-wise as follows:
\begin{equation*}
  \delta\bm{x}_k = \left[ \delta\bm{p}_k^{\brm{n}}, \delta\bm{v}_k^{\brm{n}}, \bm{\eta}_k^{\brm{n}}, \delta\bm{b}_{a,k}^{\brm{b}}, \delta\bm{b}_{\omega,k}^{\brm{b}} \right]
\end{equation*}

The $\oplus$ operator corresponds to usual vector addition for the position,
velocity, and bias states. For the orientation state, $\oplus$ is defined as
\begin{align*}
  \bm{q}_k^{\brm{nb}} &= \tilde{\bm{q}}_k^{\brm{nb}} \oplus \bm{\eta}_k^{\brm{n}} \\
  &= \exp_q{\left( \frac{\bm{\eta}_k^{\brm{n}}}{2} \right)} \odot \tilde{\bm{q}}_k^{\brm{nb}}
\end{align*}
where $\exp_q$ denotes the exponential map from $\mathfrak{so}(3)$ to
$SO(3)$~\cite{kok2017inertial}, represented as a quaternion, and $\odot$
denotes quaternion multiplication. Note that $\bm{\eta}_k^{\brm{n}}$ is
parametrized as an orientation deviation in $\brm{n}$.  A similar formulation
may be derived with the orientation deviation expressed in
$\brm{b}$~\cite{sola2017quaternion}.

The nonlinear error-state is tracked with an error-state EKF.  Owing to the
multiplicative orientation dynamics and update, this filter is sometimes
referred to as the multiplicative-EKF~\cite{crassidis2007survey}.

\subsection{State Dynamics}

Inertial measurements, collectively denoted $\bm{u}_k$, are interpreted as
control inputs during the state propagation step. The true-state dynamics function
$f_{k} \left( \bm{x}_{k}, \bm{u}_k, \bm{w}_k \right)$ is modeled as
\begin{align*}
  \bm{p}_{k+1}^{\brm{n}} &= \bm{p}_k^{\brm{n}} + T \bm{v}_k^{\brm{n}} + \frac{T^2}{2} \left( R_k^{\brm{nb}} \left( \bm{z}_{a,k}^{\brm{b}} - \bm{b}_{a,k}^{\brm{b}} - \bm{w}_{a,k}^{\brm{b}} \right) + \bm{g}^{\brm{n}} \right) \\
  \bm{v}_{k+1}^{\brm{n}} &= \bm{v}_k^{\brm{n}} + T \left( R_k^{\brm{nb}} \left( \bm{z}_{a,k}^{\brm{b}} - \bm{b}_{a,k}^{\brm{b}} - \bm{w}_{a,k}^{\brm{b}} \right) + \bm{g}^{\brm{n}} \right) \\
  \bm{q}_{k+1}^{\brm{nb}} &= \bm{q}_k^{\brm{nb}} \odot \exp_q \left( \frac{T}{2} \left( \bm{z}_{\omega,k}^{\brm{b}} - \bm{b}_{\omega,k}^{\brm{b}} - R_k^{\brm{bn}} \bm{\omega}_{\brm{e}}^{\brm{n}} - \bm{w}_{\omega,k}^{\brm{b}} \right) \right) \\
  \bm{b}_{a,k+1}^{\brm{b}} &= \bm{b}_{a,k}^{\brm{b}} + \bm{w}_{\bm{b}_a,k}^{\brm{b}} \\
  \bm{b}_{\omega,k+1}^{\brm{b}} &= \bm{b}_{\omega,k}^{\brm{b}} + \bm{w}_{\bm{b}_\omega,k}^{\brm{b}}
\end{align*}
where $T$ is the propagation duration, $R_k^{\brm{nb}}$ is the rotation matrix
representation of $\bm{q}_k^{\brm{nb}}$, $\bm{z}_{a,k}^{\brm{b}}$ and
$\bm{z}_{\omega,k}^{\brm{b}}$ are the IMU specific force and angular rate
measurements, respectively, $\bm{w}_{a,k}$ and $\bm{w}_{\omega,k}$ are the IMU
specific force and angular rate white noise, respectively, $\bm{g}^{\brm{n}}
\approx \left[ 0, 0, \SI{-9.8}{\meter\per\square\second} \right]$ is the
acceleration due to gravity after compensation for the centripetal force due to
earth's rotation, and $\bm{\omega}_{\brm{e}}^{\brm{n}}$ is the angular rate of
the earth with respect to an inertial frame.  The accelerometer and gyroscope
biases are modeled as random walk processes driven by white noise
$\bm{w}_{\bm{b}_a,k}^{\brm{b}}$ and $\bm{w}_{\bm{b}_\omega,k}^{\brm{b}}$,
respectively, whose variances are derived from the IMU bias instability
parameters~\cite{woodman2007introduction}.

The nominal-state dynamics function $f_{\mathrm{nom},k} \left( \bm{x}_{\mathrm{nom},k},
\bm{u}_k, \bm{w}_k \right)$ is similar to $f_{k} \left( \bm{x}_{k}, \bm{u}_k, \bm{w}_k \right)$:
\begin{align*}
  \tilde{\bm{p}}_{k+1}^{\brm{n}} &= \tilde{\bm{p}}_{k}^{\brm{n}} + T \tilde{\bm{v}}_k^{\brm{n}} + \frac{T^2}{2} \left( \tilde{R}_k^{\brm{nb}} \left( \bm{z}_{a,k}^{\brm{b}} - \tilde{\bm{b}}_{a,k}^{\brm{b}} \right) + \bm{g}^{\brm{n}} \right) \\
  \tilde{\bm{v}}_{k+1}^{\brm{n}} &= \tilde{\bm{v}}_{k}^{\brm{n}} + T \left( \tilde{R}_k^{\brm{nb}} \left( \bm{z}_{a,k}^{\brm{b}} - \tilde{\bm{b}}_{a,k}^{\brm{b}} \right) + \bm{g}^{\brm{n}} \right) \\
  \tilde{\bm{q}}_{k+1}^{\brm{nb}} &= \tilde{\bm{q}}_{k}^{\brm{nb}} \odot \exp_q \left( \frac{T}{2} \left( \bm{z}_{\omega,k}^{\brm{b}} - \tilde{\bm{b}}_{\omega,k}^{\brm{b}} - \tilde{R}_k^{\brm{bn}} \bm{\omega}_{\brm{e}}^{\brm{n}} \right) \right) \\
  \tilde{\bm{b}}_{a,k+1}^{\brm{b}} &= \tilde{\bm{b}}_{a,k}^{\brm{b}} \\
  \tilde{\bm{b}}_{\omega,k+1}^{\brm{b}} &= \tilde{\bm{b}}_{\omega,k}^{\brm{b}}
\end{align*}

The error-state dynamics function $f_{\mathrm{err},k} \left( \delta\bm{x}_k, \bm{u}_k,
\bm{w}_k \right)$, is straightforwardly defined as
\begin{equation*}
  f_{\mathrm{err},k}  \triangleq f_{k} \ominus f_{\mathrm{nom},k}
\end{equation*}
where $\ominus$ denotes a generalized subtraction operator similar to $\oplus$
defined earlier.

The linearized covariance propagation step of the EKF requires computation of
the following Jacobians.
\begin{align}
  F_k &= \frac{\partial \bm{f}_{\mathrm{err},k} \left( \delta\bm{x}_k, \bm{u}_k, \bm{w}_k \right)}{\partial \delta\bm{x}_k} \Bigr|_{\substack{\delta\bm{x}_k=0 \\ \bm{w}_k=0}} \label{eq:Fk} \\
  G_k &= \frac{\partial \bm{f}_{\mathrm{err},k} \left( \delta\bm{x}_k, \bm{u}_k, \bm{w}_k \right)}{\partial \bm{w}_k} \Bigr|_{\substack{\delta\bm{x}_k=0 \\ \bm{w}_k=0}} \label{eq:Gk}
\end{align}

This involves calculus of rotations. The interested reader is referred
to~\cite{sola2017quaternion, kok2017inertial} for further details. The
nontrivial sub-blocks of $F_k$ and $G_k$ are documented in
Appendix~\ref{app:jacobians}.

\subsection{Measurement Models \& Calibration}

This section details the measurement models for the various measurements
applied to the error-state EKF, along with the calibration procedures necessary
for the application of these measurements.

\subsubsection{Inertial Measurements}
\label{sec:inertial}

IMUs measure the specific force and angular rate experienced by $\brm{b}$
relative to an inertial frame. If the centripetal force due to earth's rotation
is absorbed in $\bm{g}^{\brm{n}}$, then the accelerometer and gyroscope
measurements $\bm{z}_{a,k}^{\brm{b}}$ and $\bm{z}_{\omega,k}^{\brm{b}}$,
respectively, are modeled as
\begin{align*}
  \bm{z}_{a,k}^{\brm{b}} &= R_k^{\brm{bn}} \left( \bm{a}_k^{\brm{n}} - \bm{g}^{\brm{n}} \right) + \bm{b}_{a,k}^{\brm{b}} + \bm{w}_{a,k}^{\brm{b}} \\
  \bm{z}_{\omega,k}^{\brm{b}} &= \bm{\omega}_k^{\brm{b}} + R_k^{\brm{bn}} \bm{\omega}_{\brm{e}}^{\brm{n}} + \bm{b}_{\omega,k}^{\brm{b}} + \bm{w}_{\omega,k}^{\brm{b}}
\end{align*}
where $\bm{a}_k^{\brm{n}}$ is the true acceleration of the IMU in the $\brm{n}$
frame, which double-integrates to position deviation, and
$\bm{\omega}_k^{\brm{b}}$ is the true angular rate of the IMU in the $\brm{n}$
frame, which integrates to orientation deviation. For low-quality IMUs,
accelerometer and gyroscope scale factors may also need to be modeled. For the
MEMS IMU used in this work, it was observed that modeling the scale factors did
not yield any performance benefit.

The stochastic models for IMU white noise and random walk process are derived
from the IMU specifications. In addition to such intrinsic calibration,
extrinsic calibration of the IMU with respect to $\brm{s}$ is necessary for the
application of other measurements expressed in $\brm{s}$. The vector
$\bm{p}_{\brm{sb}}^{\brm{s}}$ from $\brm{s}$ to $\brm{b}$ is taken to be known
from the mechanical specification since this is not strongly observable from
the available measurements. It is, however, important to estimate any
deviations from the mechanically specified orientation
$\bar{\bm{q}}^{\brm{sb}}$ between $\brm{b}$ and $\brm{s}$, since even
sub-degree errors in the IMU orientation relative to $\brm{s}$ may lead to
substantial errors when multiplied with the lever arm to another sensor.

The orientation deviation of $\bar{\bm{q}}^{\brm{sb}}$ from truth, denoted
$\bm{\eta}_{\brm{sb}}^{\brm{s}}$, can be effectively estimated when CDGNSS
measurements from multiple antennas are available to the EKF, as will be
discussed in Sec.~\ref{sec:cdgnss}. Accordingly, the state vector
$\delta\bm{x}_k$ is augmented with $\bm{\eta}_{\brm{sb}}^{\brm{s}}$ during
clear-sky periods.  It must be noted, however, that since the IMU is mounted
near the line connecting the Sensorium's two GNSS antennas, only two of the
three elements in $\bm{\eta}_{\brm{sb}}^{\brm{s}}$ are strongly observable. Any
orientation deviation about the vector joining the two antennas is poorly
unobservable, and must be constrained by construction. Also note that
estimation of $\bm{\eta}_{\brm{sb}}^{\brm{s}}$ only need be performed once as
long as all sensors are rigidly mounted, and may not even be necessary if the
mechanical tolerances are acceptably small.

\subsubsection{CDGNSS Measurements}
\label{sec:cdgnss}

CDGNSS offers cm-accurate position measurements under all weather conditions,
but typically offers reduced solution availability in deep urban environments.
This paper takes the approach of incorporating CDGNSS measurements in the
localization engine whenever they are available, while being capable of
maintaining the required lane-level accuracy over long CDGNSS outages in deep
urban canyons. In essence, the approach developed in this paper leverages
CDGNSS for periodic or one-time intrinsic and extrinsic calibration of other
on-board sensors, and relies on these sensors for accurate localization when
CDGNSS is unavailable.

Signals captured from the two GNSS antennas on the Sensorium are processed
together with those from a nearby reference station to provide nearly-independent
three-dimensional position measurements of the antennas in the $\brm{n}$ frame.
The position measurement for antenna $\brm{a}_i,~i \in \{0,1\}$ is
modeled as
\begin{equation}
  \bm{z}_{\brm{a}_i, k}^{\brm{n}} = \bm{p}_k^{\brm{n}} + R_k^{\brm{nb}} R^{\brm{bs}} \bm{p}_{\brm{ba}_i}^{\brm{s}} + \bm{e}_{\brm{a}_i, k}
  \label{eq:cdgnss}
\end{equation}
where $\bm{e}_{\brm{a}_i, k}$ is the CDGNSS measurement noise. The vector
$\bm{p}_{\brm{ba}_i}^{\brm{s}}$ from $\brm{b}$ to the antenna $\brm{a}_i$,
expressed in $\brm{s}$, is available from the mechanical specification. As
discussed above, $R^{\brm{bs}}$ may be taken to be the same as
$\bar{R}^{\brm{bs}}$ from the mechanical specification, or may be further
calibrated by augmenting the state with $\bm{\eta}_{\brm{sb}}^{\brm{s}}$.

Additionally, the error-state EKF requires the Jacobian of the measurement
model with respect to the error state:
\begin{equation*}
  H_{\brm{a}_i, k} \triangleq
      \frac{\partial \bm{z}_{\brm{a}_i, k}^{\brm{n}}}{\partial \delta\bm{x}_k} \Bigr|_{\substack{\delta\bm{x}_k=0 \\ \bm{e}_{\brm{a}_i, k}=0}}
      = \frac{\partial \bm{z}_{\brm{a}_i, k}^{\brm{n}}}{\partial \bm{x}_k} \Bigr|_{\substack{\bm{x}_k=\bm{x}_{\mathrm{nom},k} \\ \bm{e}_{\brm{a}_i, k}=0}} \cdot
        \frac{\partial \bm{x}_k}{\partial \delta\bm{x}_k} \Bigr|_{\substack{\delta\bm{x}_k=0 \\ \bm{e}_{\brm{a}_i, k}=0}}
\end{equation*}

The nontrivial sub-blocks of $H_{\brm{a}_i, k}$ are documented in
Appendix~\ref{app:jacobians}.

\subsubsection{Radar Range Rate \& Bearing Measurements}

The range rate and bearing measurements from automotive radars provide a
valuable velocity constraint for inertial navigation. Importantly, the
frequency modulated continuous wave (FMCW) signal used in automotive radars
provides instantaneous range rate measurements to the detected targets, i.e.,
target tracking and/or matching across cluttered radar scans is not necessary
to obtain and apply this measurement.

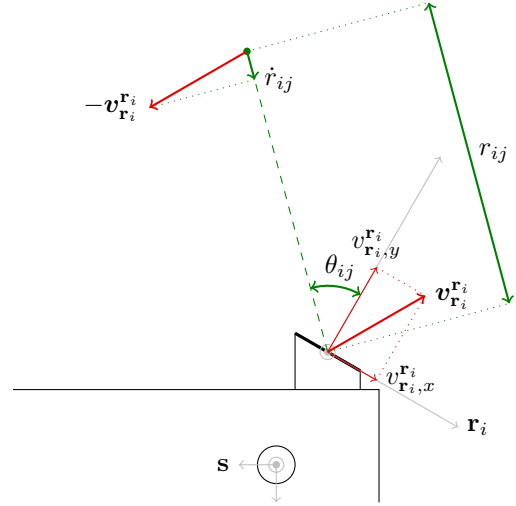
\begin{figure}[ht]
  \centering
  \begin{tikzpicture}[scale=.5]

  \coordinate(O) at (0,0);

  \path (O) +(-30:1) coordinate (r3);
  \path (O) +(150:1) coordinate (r2);
  \path (r2) +(0,{-2*sin(30)-0.5}) coordinate (r1);
  \path (r3) +(0,-0.5) coordinate (r4);
  \draw (r1) -- (r2);
  \draw[very thick] (r2) -- (r3);
  \draw (r3) -- (r4);

  \path (r1) +(-7.5,0) coordinate (s1);
  \path (r4) +(0.5,0) coordinate (s2);
  \path (s2) +(0,-3) coordinate (s3);
  \draw (s1) -- (s2) -- (s3);

  \path (r1) +(-0.5,-2.0) coordinate (A);
  \draw (A) circle [radius=0.5];
  \draw[->, black!25] (A) -- ++(-1,0) node[anchor=east, black] {$\brm{s}$};
  \draw[->, black!25] (A) -- ++(0,-1);
  \draw[black!25] (A) circle [radius=0.2];
	\fill[black!25] (A) circle [radius=0.1];

  \draw[->, black!25] (O) -- ++(-30:4) node[anchor=west, black] {$\brm{r}_i$};
  \draw[->, black!25] (O) -- ++(+60:6) coordinate (ry);
  \draw[black!25] (O) circle [radius=0.2];
	\fill[black!25] (O) circle [radius=0.1];

  \draw[->, black!10!red] (O) -- ++(+150:-1.5) coordinate (vx) node[anchor=west, black] {$v^{\brm{r}_i}_{{\brm{r}_i},x}$};
  \draw[->, black!10!red] (O) -- ++(+60:{sqrt(3)*1.5}) coordinate (vy) node[anchor=south, black] {$v^{\brm{r}_i}_{{\brm{r}_i},y}$};
  \draw[dotted, black!10!red] (vx) -- ++(+60:{sqrt(3)*1.5}) coordinate (vri);
  \draw[dotted, black!10!red] (vy) -- ++(+150:-1.5);
  \draw[->, thick, black!10!red] (O) -- (vri) node[anchor=west, black] {$\bm{v}^{\brm{r}_i}_{{\brm{r}_i}}$};

  \path (O) +({-8*tan(15)},8) coordinate (T);
  \fill[black!50!green] (T) circle [radius=0.1];
  \path (T) +(+150:1.5) coordinate (mvx);
  \path (mvx) +(-120:{sqrt(3)*1.5}) coordinate (mvri);
  \draw[->, thick, black!10!red] (T) -- (mvri) node[anchor=east, black] {$-\bm{v}^{\brm{r}_i}_{{\brm{r}_i}}$};
  \draw[dashed, black!50!green] (O) -- (T);
  \draw[->, thick, black!50!green] (T) -- ++(-75:{1.5*cos(75)/sin(30)}) coordinate (vrdot) node[anchor=west, black] {$\dot{r}_{ij}$};
  \draw[dotted, black!50!green] (mvri) -- (vrdot);

  \pic["$\theta_{ij}$", draw=black!50!green, thick, <->, angle radius=25, pic text options={shift={(0.1,0.6)}}] {angle=ry--O--T};

  \draw[dotted, black!50!green] (O) -- ++(+15:5) coordinate (rO);
  \draw[dotted, black!50!green] (T) -- ++(+15:5) coordinate (rT);
  \draw[<->, thick, black!50!green] (rO) -- node[anchor=west, black] {$r_{ij}$} (rT);

\end{tikzpicture}
  \caption{A visual description of the radar range rate measurement model.
    Quantities labeled in green are measured by the radar.  The relative
    velocity of a stationary target with respect to $\brm{r}_i$ is the negative
    of the velocity with respect to $\brm{n}$ of the $i$th radar, expressed in
    $\brm{r}_i$, written $-\bm{v}_{\brm{r}_i,k}^{\brm{r}_i}$. The measured
    radial velocity $\dot{r}_{ij}$ of the $j$th stationary target is the
    projection of $-\bm{v}_{\brm{r}_i,k}^{\brm{r}_i}$ onto the line-of-sight
    direction between the $i$th radar and the $j$th target.}
  \label{fig:radar-rdot}
\end{figure}

The relative velocity of a stationary target with respect to $\brm{r}_i$ is
given by the negative of the velocity with respect to $\brm{n}$ of the $i$th
radar, expressed in $\brm{r}_i$, written $-\bm{v}_{\brm{r}_i,k}^{\brm{r}_i}$,
as shown in Fig.~\ref{fig:radar-rdot}.  Assuming that the radar only detects
targets in the two-dimensional plane of the linear phased array, the range rate
measurement is modeled as
\begin{equation}
  \dot{r}_{ij,k} = {\begin{bmatrix} ~~\sin{\theta_{ij,k}} \\ -\cos{\theta_{ij,k}} \\ 0 \end{bmatrix}}^\top R^{\brm{r}_i \brm{s}} R^{\brm{sb}}
    \left( R_k^{\brm{bn}} \bm{v}_k^{\brm{n}} + \left( \bm{\omega}_k^{\brm{b}} \times R^{\brm{bs}} \bm{p}_{\brm{br}_i}^{\brm{s}} \right) \right)
  \label{eq:radar-rdot}
\end{equation}
where the vector $\bm{p}_{\brm{br}_i}^{\brm{s}}$ and the radar orientation
$R^{\brm{r}_i \brm{s}}$ may be taken from the mechanical specifications. Note that
unlike typical measurement models where the right-hand side is composed of
quantities that are either known or are being estimated, ~\ref{eq:radar-rdot}
has \emph{measured} quantities $\theta_{ij,k}$ on the right-hand side of the
equation. This implies that any errors in the bearing measurements will not be
accounted for if the range rate measurements are modeled in the EKF as shown.

\begin{figure}[ht]
  \centering
  \includegraphics[width=\linewidth]{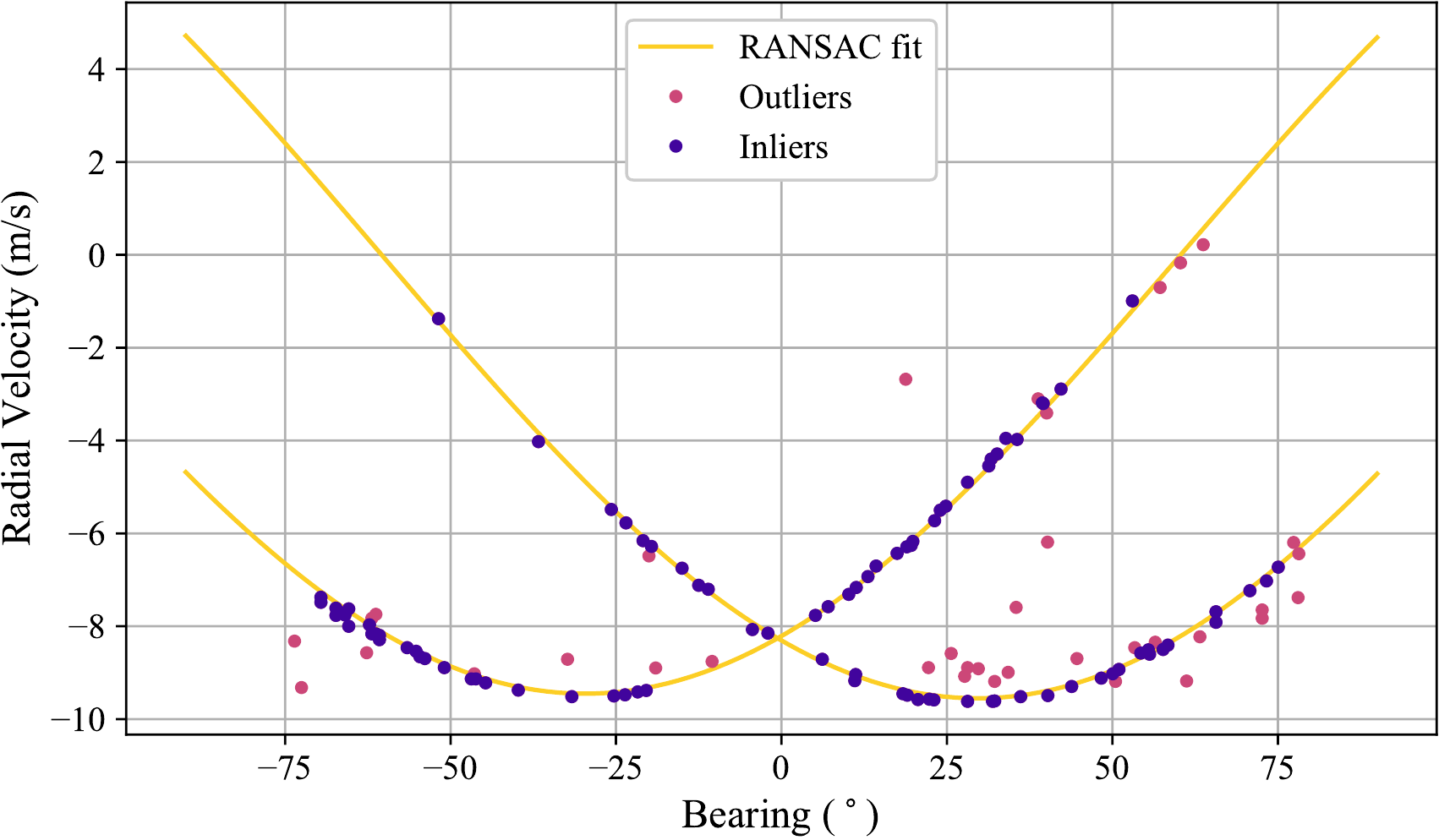}
  \caption{Example results of the RANSAC operation on radar range rate and bearing
  measurements. The two yellow sinusoidal curves represent the RANSAC-predicted
  radial velocities for the port and starboard radars from
  Fig.~\ref{fig:sensorium} as a function of the bearing.  With a threshold of
  \SI{0.2}{\meter\per\second}, RANSAC considers violet dots as inliers and
  magenta dots as outliers.  Note that the radial velocity magnitude is maximized at
  \SI{-30}{\degree} and \SI{+30}{\degree} for the port and starboard radars,
  respectively, in agreement with the mounting angles of these radars on the
  vehicle.}
  \label{fig:radar-ransac}
\end{figure}

The application of range rate constraints comes with two major challenges.
First, individual radar scans contain a number of spurious targets as discussed
in Sec.~\ref{sec:introduction}. Second, automotive phased-array radars exhibit
poor bearing resolution and accuracy, and this is further exacerbated by the
unusual range rate measurement model described above. Both of these
challenges are addressed by pre-processing the range rate and bearing
measurements with a RANSAC routine that estimates a best-fit two-dimensional
radar velocity model to the radar measurements. In particular, with $N$
detected targets, the RANSAC operation finds a robust solution to the following
system of equations:
\begin{equation}
  \begin{bmatrix} \dot{r}_{i0} \\ \vdots \\ \dot{r}_{iN} \end{bmatrix} =
    \begin{bmatrix} \sin{\theta_{i0}} & -\cos{\theta_{i0}} \\ \vdots & \vdots \\ \sin{\theta_{iN}} & -\cos{\theta_{iN}} \end{bmatrix}
    \begin{bmatrix} v_{\brm{r}_i,x}^{\brm{r}_i} \\ v_{\brm{r}_i,y}^{\brm{r}_i} \end{bmatrix}
  \label{eq:ransac}
\end{equation}
while eliminating the $\left( \dot{r}_{ij}, \theta_{ij} \right)$ pairs that may
be outliers. Example results from  the RANSAC procedure are shown in
Fig.~\ref{fig:radar-ransac}. Ultimately, the solution to ~\ref{eq:ransac} is
applied as a measurement to the EKF with the following measurement model:
\begin{align*}
  \bm{z}_{\brm{r}_i,k}^{\brm{r}_i} &\triangleq {\begin{bmatrix} v_{\brm{r}_i,x}^{\brm{r}_i} \\ v_{\brm{r}_i,y}^{\brm{r}_i} \end{bmatrix}}_k \\
  &= {\left[
        R^{\brm{r}_i \brm{s}} R^{\brm{sb}}
        \left( R_k^{\brm{bn}} \bm{v}_k^{\brm{n}} + \left( \bm{\omega}_k^{\brm{b}} \times R^{\brm{bs}} \bm{p}_{\brm{br}_i}^{\brm{s}} \right) \right)
      \right]}_{\left[ 0,1 \right]} + \bm{e}_{\brm{r}_i,k}
\end{align*}
where the subscript $\left[ 0,1 \right]$ denotes the first two elements of the
three-element vector. Parts of the Jacobian of this measurement model with
respect to the EKF error-state are documented in Appendix~\ref{app:jacobians}.

\subsubsection{Ground Vehicle Dynamics Constraints}
\label{sec:vehicle-calib}

Under nominal driving conditions, a ground vehicle respects dynamical
constraints which can be leveraged as measurements to the EKF. This paper
incorporates near-zero sideslip and vertical velocity constraints, commonly
referred to as nonholonomic constraints (NHC), as well as zero-speed updates
(ZUPT). The measurement models for these constraints are described below.

\paragraph{Nonholonomic Constraints (NHC)}

The application of NHC is based on the following assumptions:
\begin{enumerate}
  \item There exists a fixed center of rotation, taken to be the origin of
    $\brm{v}$, about which the vehicle rotates when a steering control input is
    applied.
  \item When a zero steering input is applied, the vehicle only moves in the
    $\brm{v}_y$ direction. This holds by definition of $\brm{v}$.
  \item The vehicle does not slip sideways or leave the surface of the road.
\end{enumerate}

When the above assumptions hold, it follows that the velocity of the vehicle,
when expressed in $\brm{v}$, is zero in the $\brm{v}_x$ and $\brm{v}_z$
directions at all times. In practice, however, these assumptions only hold
approximately. Accordingly, the zero sideslip and vertical velocity constraints
are applied as \emph{soft} constraints in the form of measurements with an
associated measurement error covariance. The NHC is modeled as
\begin{align}
  \bm{0}_{2 \times 1}
  &\triangleq \bm{z}_{\mathrm{nhc},k}^{\brm{v}} \\
  &= {\left[ \bm{v}_k^{\brm{v}} \right]}_{\left[ 0,2 \right]} + \bm{e}_{\mathrm{nhc},k} \notag \\
  &= {\left[
        R^{\brm{vs}} R^{\brm{sb}} \left( R_k^{\brm{bn}} \bm{v}_k^{\brm{n}} + \left( \bm{\omega}_k^{\brm{b}} \times R^{\brm{bs}} \bm{p}_{\brm{bv}}^{\brm{s}} \right) \right)
      \right]}_{\left[ 0,2 \right]} + \bm{e}_{\mathrm{nhc},k} \label{eq:nhc}
\end{align}
where
$\bm{p}_{\brm{bv}}^{\brm{s}} = \bm{p}_{\brm{bs}}^{\brm{s}} +
\bm{p}_{\brm{sv}}^{\brm{s}}$ and $R^{\brm{vs}}$ are parts of the extrinsic
calibration between $\brm{v}$ and $\brm{s}$. Precise manual measurement of
$\bm{p}_{\brm{sv}}^{\brm{s}}$ and $R^{\brm{vs}}$ is challenging. First, it is
not obvious where the origin of $\brm{v}$ lies, though the center of line
connecting the two rear axles might be a reasonable guess.  Second, it would be
challenging to measure, for example, the pitch of the Sensorium relative to the
plane of the vehicle chassis. Accordingly, this paper takes a data-driven
approach to extrinsic calibration of $\brm{v}$.

Once again, the extrinsic calibration technique relies on clear-sky periods
with good CDGNSS availability, such that the nominal state estimates of
$\bm{v}_k^{\brm{n}}$, $\bm{q}_k^{\brm{nb}}$, and $\bm{b}_{\omega,k}^{\brm{b}}$
are close to their true values. Furthermore, calibration begins with coarse
initial guesses of $R^{\brm{vs}}$ and $\bm{p}_{\brm{sv}}^{\brm{s}}$, denoted
$\bar{R}^{\brm{vs}}$ and $\bar{\bm{p}}_{\brm{sv}}^{\brm{s}}$, respectively, and
attempts to estimate the orientation deviation $\bm{\eta}_{\brm{vs}}^{\brm{s}}$
and lever arm deviation $\delta\bm{p}_{\brm{sv}}^{\brm{s}}$ with respect to
these. With other quantities assumed known, ~\ref{eq:nhc} may be rewritten as
\begin{align*}
  \bm{e}_{\mathrm{nhc},k}
  &= {\left[
        \left( \bar{R}^{\brm{vs}} \oplus \bm{\eta}_{\brm{vs}}^{\brm{s}} \right)
        \left( \bm{v}_k^{\brm{s}} + \left( \bm{\omega}_k^{\brm{s}} \times \left( \bar{\bm{p}}_{\brm{bv}}^{\brm{s}} + \delta\bm{p}_{\brm{bv}}^{\brm{s}} \right) \right) \right)
      \right]}_{\left[ 0,2 \right]} \\
  &\triangleq \bm{h}_{\mathrm{nhc},k} \left( \bm{\eta}_{\brm{vs}}^{\brm{s}}, \delta\bm{p}_{\brm{bv}}^{\brm{s}} \right)
\end{align*}

This model is nonlinear in $\bm{\eta}_{\brm{vs}}^{\brm{s}}$, and may be solved
as a nonlinear least squares problem, e.g., with the Gauss-Newton method. The
Jacobian of $\bm{h}_{\mathrm{nhc},k}$ evaluated at
$\bm{\eta}_{\brm{vs}}^{\brm{s}} = \bm{0}$ and
$\delta\bm{p}_{\brm{bv}}^{\brm{s}} = \bm{0}$ is composed of
\begin{align*}
  \frac{\partial \bm{h}_{\mathrm{nhc},k}}{\partial \bm{\eta}_{\brm{vs}}^{\brm{s}}}
  &= \left[ {\left( \bm{v}_k^{\brm{s}} + \bm{\omega}_k^{\brm{s}} \times \bar{\bm{p}}_{\brm{bv}}^{\brm{s}} \right)}^\top
            \otimes {\left[ \bar{R}^{\brm{vs}} \right]}_{\left[ (0,2),(:) \right]}
     \right]
     \begin{bmatrix} {[ -\hat{\brm{i}} ]}_{\times} \\ {[ -\hat{\brm{j}} ]}_{\times} \\ {[ -\hat{\brm{k}} ]}_{\times} \end{bmatrix} \\
  \frac{\partial \bm{h}_{\mathrm{nhc},k}}{\partial \delta\bm{p}_{\brm{bv}}^{\brm{s}}}
  &= {\left[ \bar{R}^{\brm{vs}} \right]}_{\left[ (0,2),(:) \right]} {\left[ \bm{\omega}_k^{\brm{s}} \right]}_{\times}
\end{align*}
where $\otimes$ denotes the Kronecker product, subscript $\left[ (0,2),(:)
\right]$ denotes selection of the first and third rows of a matrix, ${[ \cdot
]}_{\times}$ denotes the skew-symmetric cross-product matrix corresponding to
the \num{3}-element argument, and $\hat{\brm{i}}$, $\hat{\brm{j}}$, and
$\hat{\brm{k}}$ denote the cardinal unit vectors. To make the system
observable, measurements from multiple epochs must be stacked and solved as a
batch.  Additionally, the nonlinear problem must be iteratively linearized and
solved until convergence.

\paragraph{Zero-Speed Update (ZUPT)}

The ZUPT constraint is another valuable measurement that limits odometric
drift, especially in situations where the platform makes frequent stops. The
measurement model for ZUPT is trivially written as
\begin{align}
  \bm{0}_{3 \times 1} &\triangleq \bm{z}_{\mathrm{zupt},k}^{\brm{v}} \notag \\
  &= R^{\brm{vs}} R^{\brm{sb}} R_k^{\brm{bn}} \bm{v}_k^{\brm{n}} + \bm{e}_{\mathrm{zupt},k} \label{eq:zupt}
\end{align}

The primary challenge of applying ZUPT is detection of epochs where
this constraint is valid. Importantly, this condition must be detected
independently from the EKF state estimate, e.g., by inspection of the raw IMU
measurements. In theory, it is not possible to make any claims about zero speed
based on acceleration and/or angular rate data, since IMU measurements of a
vehicle moving with a constant velocity and orientation must be
indistinguishable from those of a stationary vehicle.  In practice, however,
the IMU measurements exhibit a distinct behavior when the vehicle is in motion,
e.g., due to road roughness and vehicle vibrations. Prior work has made use of
these \emph{artifacts} to detect stationary periods. This paper follows the
angular rate energy method from~\cite{skog2010zero} for ZUPT detection. In
practice, if wheel odometry data are available from the vehicle CAN bus, as is
common in most modern vehicles, then ZUPT detection can be performed trivially
and with high reliability.

An observant reader might wonder why ZUPT is not applied directly to
$\bm{v}_k^{\brm{n}}$ in ~\ref{eq:zupt}. The advantage of applying ZUPT in
$\brm{v}$ is that a tighter zero-speed constraint can be reliably applied in
the lateral and vertical directions.

\subsection{Batch Smoothing \& Update}

Real-time estimates of the vehicle pose trajectory obtained from the EKF may be
used to string together individual scans and perform a radar batch measurement
update. However, since these data are processed batches, it is desirable to
perform backward smoothing over the short duration of the batch. Backward
smoothing enforces the dynamics function backwards in time, ironing out any
large jumps that may have occurred in the EKF forward pass.

Accordingly, the batch smoother component in Fig.~\ref{fig:fusion-arch} stacks
all inertial measurements and snapshots of the estimator state over the
duration of the batch. When the batch is ready to be processed for correlation,
backward smoothing is enforced with the inertial measurements as control
inputs. The smoothing formulation in this case is somewhat more complicated than
usual~\cite{sarkka2013bayesian} due to nonlinear backward dynamics and the
error-state formulation. Details on nonlinear error-state Rauch-Tung-Striebel
smoothing are provided in Appendix~\ref{app:smoothing}.

The correlation peak search region is taken to be \SI{+-5}{\meter} and
\SI{+-3}{\degree}. The 3-DoF pose offset $\widehat{\bm{\Theta}}$ from radar
batch correlation is applied as horizontal position and heading measurements to
the EKF. Outliers from batch correlation are excluded in the EKF based on a
$\chi^2$-test on the normalized innovation squared
(NIS)~\cite{y_barshalom01_tan}.

\section{Experimental Results}
\label{sec:results}

The radar-inertial positioning system of Fig.~\ref{fig:fusion-arch} was
evaluated experimentally using the dataset described
in~\cite{narula2020texcup}, collected during approximately \SI{1.5}{\hour} of
driving on two separate days in and around the urban center of Austin, TX. This
section presents the evaluation results.

\subsection{Dataset}

\begin{figure}[ht]
  \centering
  \includegraphics[width=\linewidth,trim={300 0 300 0},clip]{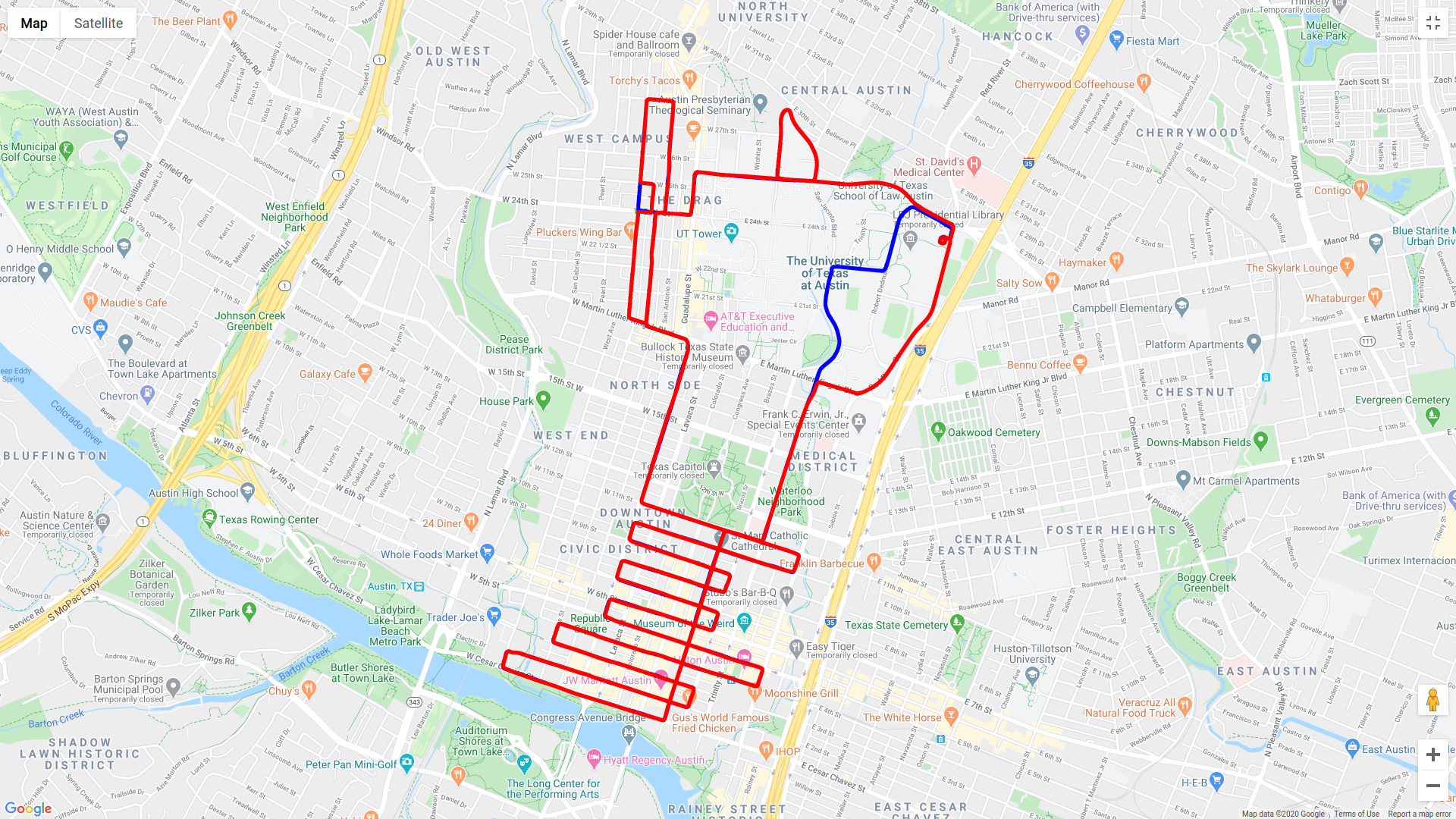}
  \caption{Test route through The University of Texas west campus and Austin
    downtown. These areas are the most challenging for precise GNSS-based
    positioning and thus would benefit the most from radar-based
    positioning. The route was driven once on a weekday and again on the
    weekend to evaluate robustness of the radar map to changes in traffic and
    parking patterns. Red is the mapping run (May 12), blue is the localization
    run (May 9).  A prior map is not available in the visible blue areas.}
  \label{fig:deep-urban-route}
\end{figure}

Fig.~\ref{fig:deep-urban-route} shows the route followed by the
sensor-instrumented vehicle on Thursday, May 9, 2019 (in blue) and Sunday, May
12, 2019 (in red). The test route combs through every street in the Austin, TX
downtown area, since such environments are the most challenging for
CDGNSS-based positioning~\cite{humphreys2019deepUrbanIts} and would benefit the
most from multi-sensor all-weather positioning.  The route was driven once on a
weekday and again on the weekend to evaluate robustness of the proposed
map-based approach to changes in the traffic and parking patterns. Note that
the final part of the route (the north-east segment) was different on the two
days, preventing the use of a map-based positioning approach. This section of
the test route has been omitted from the evaluation results.

\subsubsection{Sensors}

The Sensorium, shown in Fig.~\ref{fig:sensorium}, features two types of
automotive radars: one Delphi electronically-scanning radar (ESR) in the middle
and two Delphi short-range radars (SRR2s) on the two sides. Both the ESR and
the SRR2 are commercially available; similar radars are available on
economy-class consumer vehicles. The ESR provides simultaneous sensing in a
narrow (\SI{+-10}{\degree}) long-range (\SI{175}{\meter}) coverage area and a
wider (\SI{+-45}{\degree}) medium-range (\SI{60}{\meter}) area. The SRR2 units
each have a coverage area of \SI{+-75}{\degree} and \SI{80}{\meter}
(see~\cite[Fig.~6]{narula2020radarpositioning}).  Each SRR2 is installed facing
outward from the center-line at an angle of \SI{30}{\degree}.  The Sensorium's
onboard computer timestamps and logs the radar returns from the three radar
units.

The LORD MicroStrain 3DM-GX5-25 MEMS IMU is an industrial-grade inertial sensor
that acts as the core sensor of the localization pipeline. The IMU provides
temperature-compensated accelerometer and gyroscope readings at
\SI{100}{\hertz}. Two Antcom G8Ant-3A4TNB1 high performance GNSS patch antennas
pull in signals from all three GNSS frequency bands and include a
\SI{40}{\decibel} active low-noise amplifier.

\subsubsection{Ground-Truth Trajectory}

The ground-truth position and orientation trajectory for the data are generated
with the iXblue ATLANS-C, a high-performance CDGNSS coupled fiber-optic
gyroscope INS. The post-processed position solution obtained from the
ATLANS-C is decimeter-accurate throughout the dataset.

\subsubsection{Dataset Splits}

With a limited amount of field data available for development and evaluation,
it is critical to ensure that the proposed positioning technique does not
overfit this particular dataset. Accordingly, the data used in the development
of the algorithms were restricted to a fixed \SI{30}{\minute} segment, where
the prior radar map was constructed with radar measurements from May 9 and
localization was performed with radar, inertial, and CDGNSS measurements from
May 12. In contrast, during evaluation the full \SI{62}{\minute} of data were
used, and the mapping and localization datasets were inverted, i.e., the prior
map was constructed with radar measurements from May 12, and localization was
performed with all sensor data from May 9. The algorithms have not been
modified to maximize the performance over the evaluation dataset.

\subsection{Prior Radar Mapping}

The first step to radar-map-based localization is the generation of a radar map
point cloud. Radar scans collected from the May 12, 2019 drive were aggregated
to create a map with the benefit of the ATLANS-C ground-truth trajectory. In a
practical system, the radar map may be generated during favorable conditions
for optical sensors such as cameras and lidar, such that the mapping vehicle
can accurately track its pose. Additionally, the mapping process may be
crowed-sourced from consumer vehicles~\cite{narula2018accuracyLimits,
narula2018accurate}. The map point cloud is stored in a k-d tree for efficient
querying during localization.

Two implementation notes are in order here. First, automotive radar clutter is
especially intense when the vehicle is stationary.  Accordingly,
radar range measurements obtained when the vehicle was moving slower than
\SI{1}{\meter\per\second} were discarded for both mapping and localization. This
implies that radar correlation measurements were only available during periods
when the vehicle was moving faster than \SI{1}{\meter\per\second}. Second, it
was observed that radar returns far from the vehicle are mostly clutter and
have negligible resemblance to the surrounding structure. Radar returns with
range larger than \SI{50}{\meter} were discarded for both the map and batch
PHDs. It is noted that these two parameters have not been optimized to produce
the smallest estimation errors; instead they have been fixed based on visual
inspection.

\subsection{Offline Calibration}

Extrinsic calibration among the IMU frame $\brm{b}$, the Sensorium frame
$\brm{s}$, and the vehicle frame $\brm{v}$ was performed offline with
\SI{125}{\second} of sensor data with CDGNSS availability. While it is possible
to estimate the calibration parameters online, it may not be desirable to do so
if these parameters are not expected to change over time.

The orientation deviation $\bm{\eta}_{\brm{sb}}^{\brm{s}}$ between the IMU body
frame and the Sensorium frame was calibrated for the localization dataset, as
described in Sec.~\ref{sec:inertial}. With two GNSS antennas, only two out of
the three DoFs in $\bm{\eta}_{\brm{sb}}^{\brm{s}}$ are observable.
Accordingly, the orientation deviation around $\brm{b}_x$, which is mostly
unobservable, was tightly constrained to the initial guess of zero. The
deviations around $\brm{b}_y$ and $\brm{b}_z$ rapidly converged to sub-degree
offsets from the mechanical specification.

Extrinsic calibration between $\brm{v}$ and $\brm{s}$ was similarly estimated
over the \SI{125}{\second} period as detailed in Sec.~\ref{sec:vehicle-calib}.

The commercial automotive radars on the Sensorium do not offer any mechanism to
synchronize their scans with an external reference clock. Analysis of the radar
range rate residuals in the EKF showed clear evidence of latency between the
data logging timestamp and the true scan times. Accordingly, radar latency
calibration was performed offline with a best fit approach.

\subsection{Implementation Notes}

A few implementation- and dataset-specific notes relating to the localization
pipeline are documented below.

\paragraph{CDGNSS Measurements \& Outages}

The CDGNSS position measurements used in this evaluation are in fact the output
of the post-processed ground-truth system, i.e., these measurements have not
been obtained from an unaided CDGNSS receiver. While this is not ideal for
realistic evaluation, the
evaluation results presented herein do not mislead because, first, CDGNSS
measurements are only applied for a \SI{125}{\second} period for initial
calibration, and second, any commercial CDGNSS receiver would be able to
generate similar cm-accurate position solutions in the clear-sky region where
the CDGNSS measurements were applied.

\paragraph{Measurement Noise Correlation}

Observations from the field data revealed that the measurement noise in the
radar range rate measurements is not independent between consecutive radar
scans. This is problematic since the EKF applied assumes each measurement to
have errors that are uncorrelated in time.  Accordingly, the radar range rate
measurements were decimated to \SI{1}{\hertz} such that the measurements were
spaced out by roughly the decorrelation time of the measurement noise. A more
principled approach to this problem is to augment the state vector with states
to pre-whiten the measurements. But this approach was empirically observed to
not outperform the straightforward measurement decimation, while introducing
additional complexity and tuning parameters.

Similarly, the NHC and ZUPT measurements can in theory be applied at every
applicable IMU epoch.  But to prevent correlated errors in these constraints
(e.g., due to sideslip experienced while cornering) from making the EKF
inconsistent, they are only applied at \SI{1}{\hertz}.

\paragraph{Filter Tuning Parameters}

The process noise covariance used in the EKF is derived from the IMU datasheet
parameters~\cite{woodman2007introduction, lordmicrostrain_3dm_gx5_25}. The
measurement noise covariance associated with CDGNSS measurements is available
directly from the ATLANS-C receiver. A few other measurement noise standard
deviations and tuning parameters are documented in
Table~\ref{tab:filter-params}.

\begin{table}[htbp]
  \centering
  \caption{A List of Parameters Involved in the Localization Pipeline}
  \begin{tabular}[c]{ll}
    \toprule
    Minimum speed for valid radar range & \SI{1}{\meter\per\second} \\
    Maximum valid radar range & \SI{50}{\meter} \\
    Minimum RANSAC inliers & \num{10} \\
    Minimum fraction of RANSAC inliers & \num{0.65} \\
    $v_{\brm{r}_i,x}^{\brm{r}_i}$ (broadside) standard deviation & \SI{0.2}{\meter\per\second} \\
    $v_{\brm{r}_i,y}^{\brm{r}_i}$ (boresight) standard deviation & \SI{0.1}{\meter\per\second} \\
    $v_{\mathrm{nhc},x}^{\brm{v}}$ (lateral) standard deviation & \SI{0.1}{\meter\per\second} \\
    $v_{\mathrm{nhc},z}^{\brm{v}}$ (vertical) standard deviation & \SI{0.2}{\meter\per\second} \\
    \bottomrule
  \end{tabular}
  \label{tab:filter-params}  
\end{table}

\subsection{Localization Results}

This section presents empirical error statistics obtained from field evaluation
of the proposed method. The test scenario evaluated in this section is an
extreme one: the vehicle starts off in a clear-sky environment with
\SI{125}{\second} of CDGNSS availability, and subsequently all CDGNSS
measurements are cut off for the next \SI{3600}{\second} of driving, during
which the system must rely on radar and inertial sensing along with vehicle
dynamical constraints to maintain an accurate estimate of its pose.

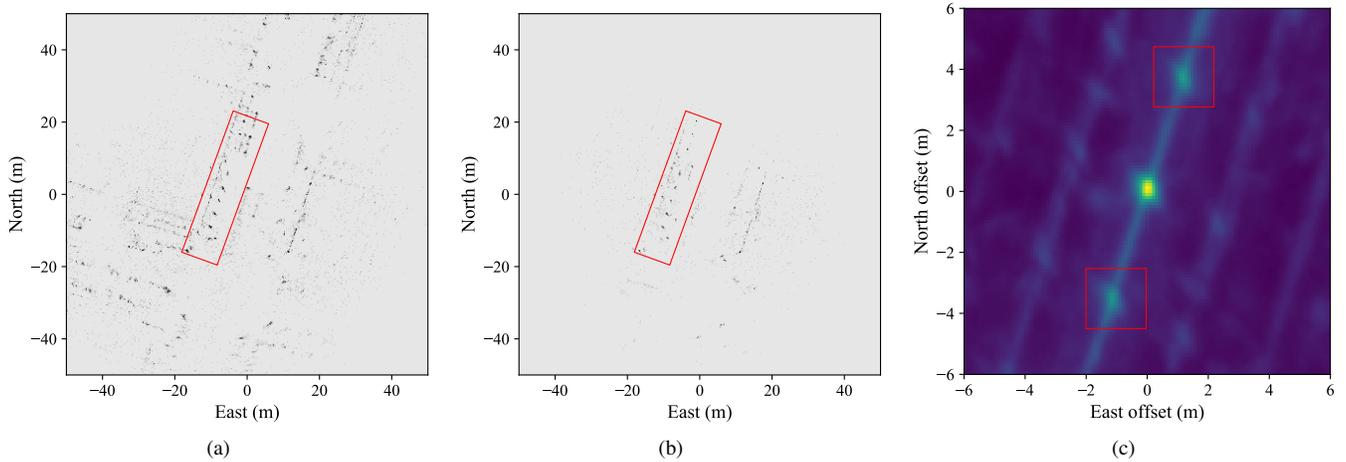
\begin{figure*}[ht]
  \centering
  \begin{minipage}[b]{0.325\textwidth}
    \centering
    \begin{tikzpicture}
      \node[inner sep=0pt] at (0,0)
      {\includegraphics[width=0.95\linewidth]{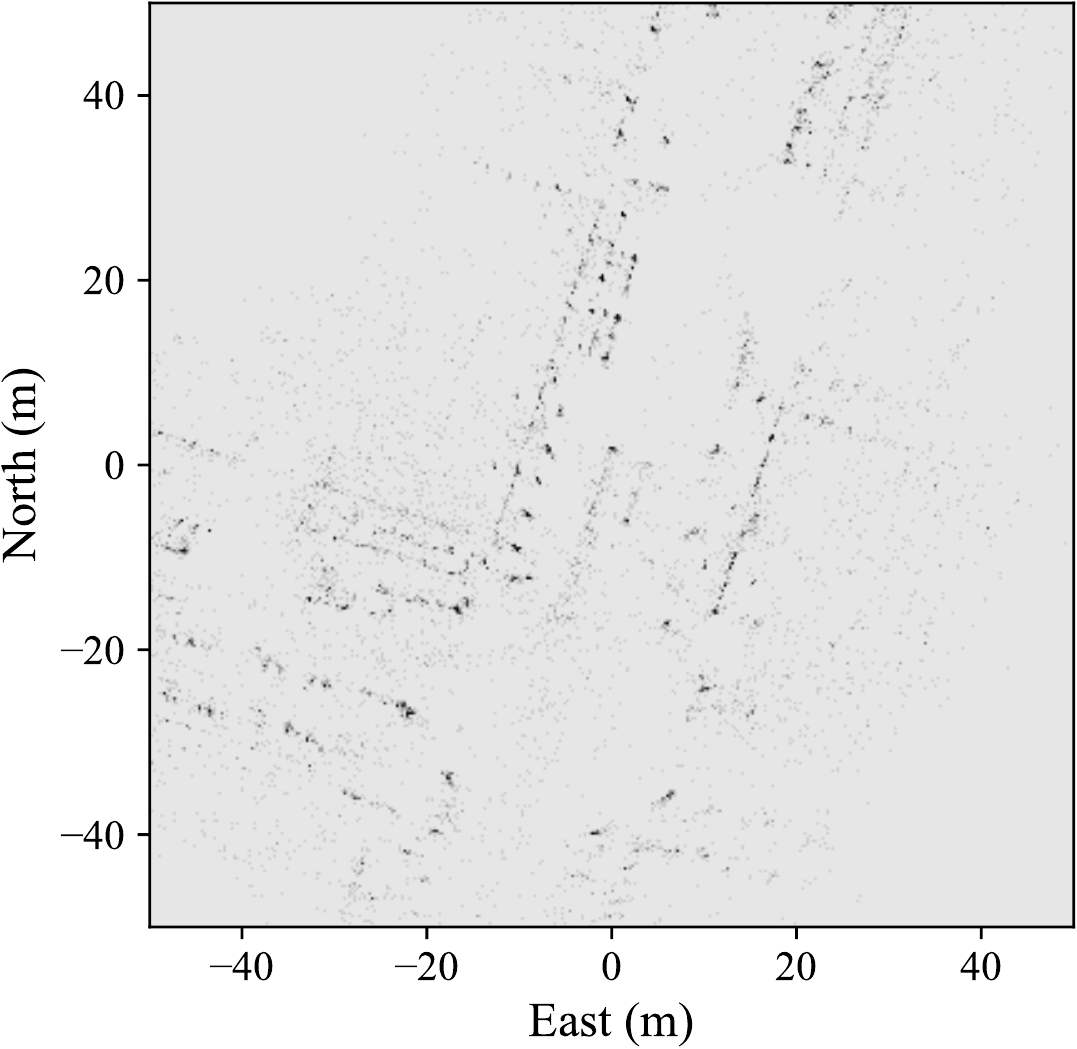}};
      \draw[color=red, rotate=-20] (-0.3,-0.6) rectangle +(0.5,2);
    \end{tikzpicture}
    \subcaption{}
    \label{fig:map-ogm}
  \end{minipage}
  \begin{minipage}[b]{0.325\textwidth}
    \centering
    \begin{tikzpicture}
      \node[inner sep=0pt] at (0,0)
      {\includegraphics[width=0.95\linewidth]{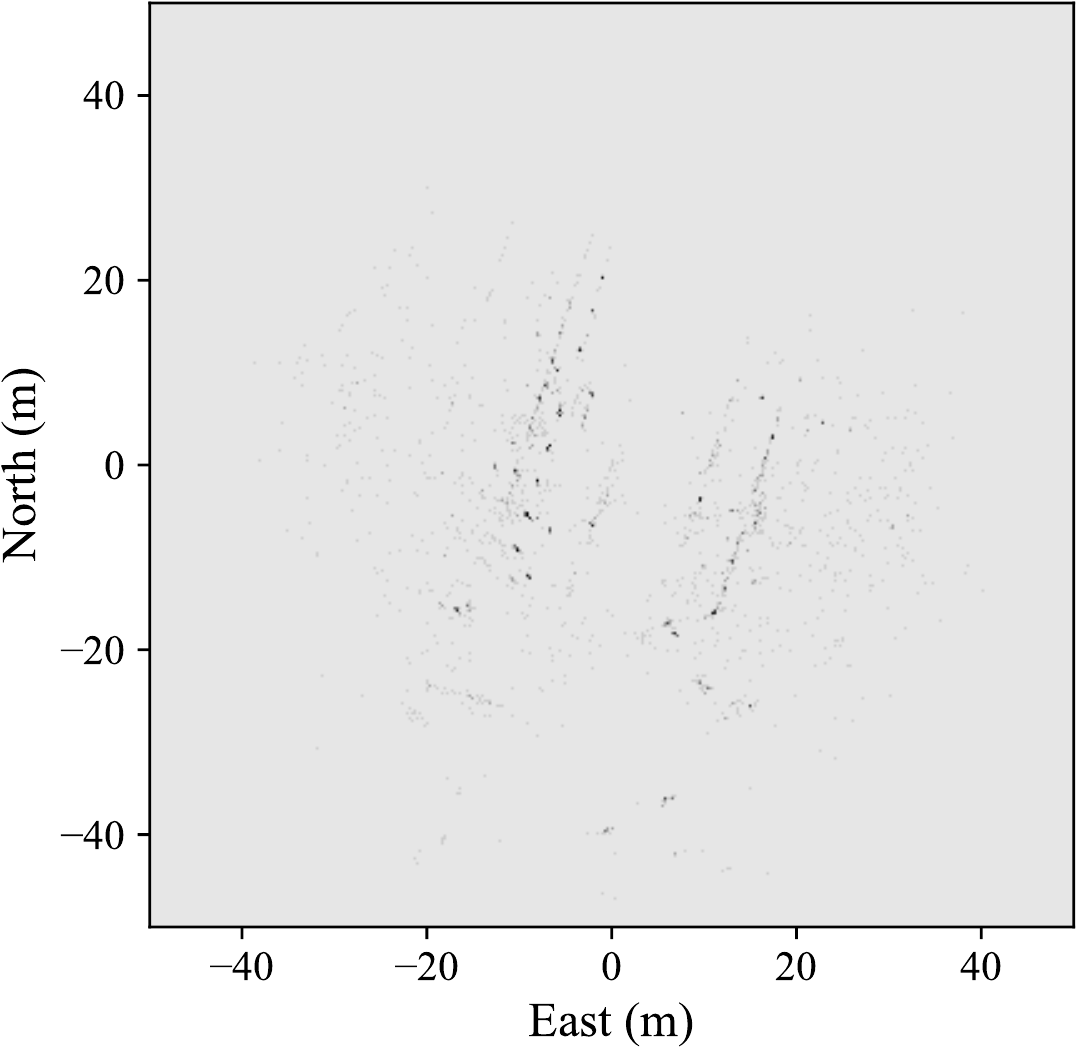}};
      \draw[color=red, rotate=-20] (-0.3,-0.6) rectangle +(0.5,2);
    \end{tikzpicture}
    \subcaption{}
    \label{fig:batch-ogm}
  \end{minipage}
  \begin{minipage}[b]{0.325\textwidth}
    \centering
    \begin{tikzpicture}
      \node[inner sep=0pt] at (0,0)
      {\includegraphics[width=0.95\linewidth]{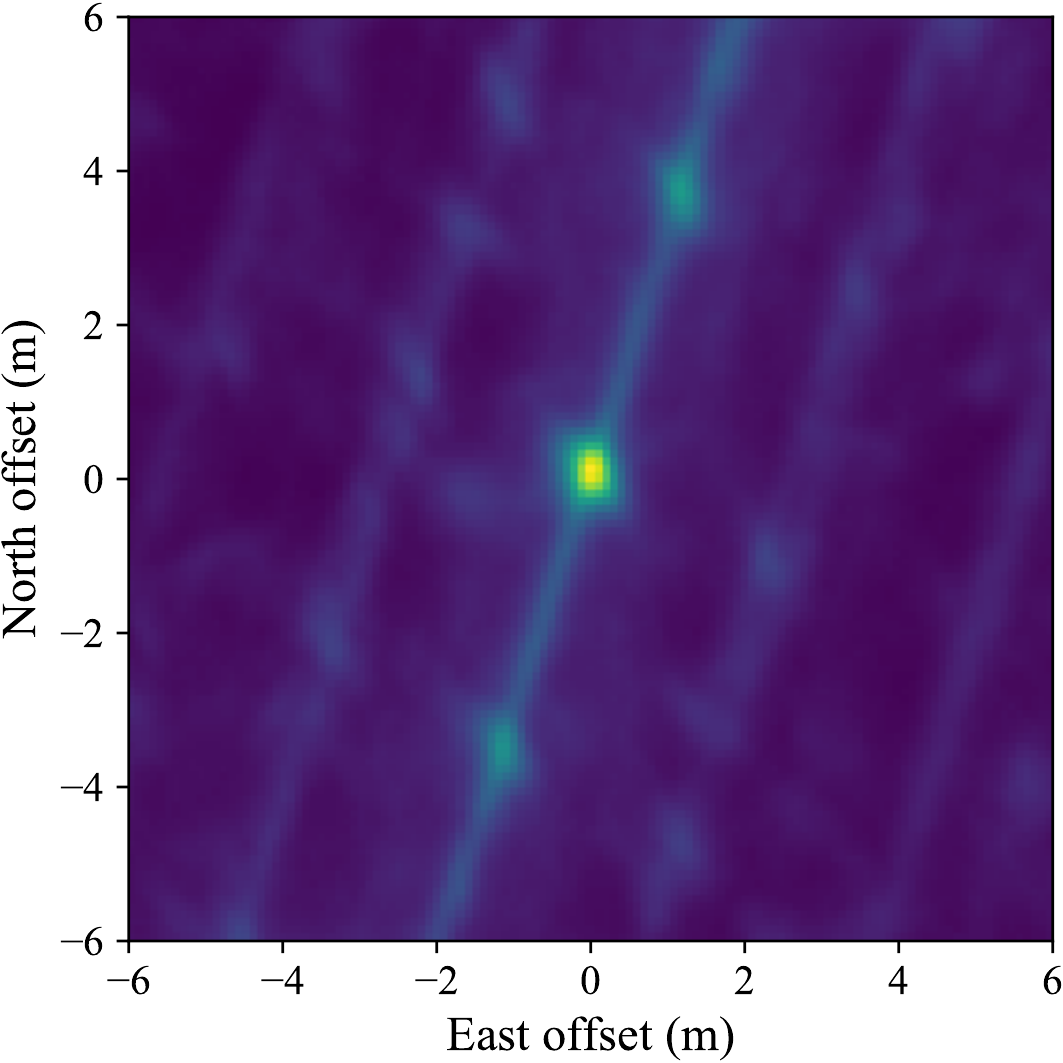}};
      \draw[color=red] (0.4,1.4) rectangle (1.2,2.2);
      \draw[color=red] (-0.5,-1.55) rectangle (0.3,-0.75);
    \end{tikzpicture}
    \subcaption{}
    \label{fig:xcorr}
  \end{minipage}
  \caption{This figure shows an interesting example of radar-based urban
  positioning with the proposed method. Panel (a) shows the occupancy grid
  estimated from the prior map point cloud. Panel (b) shows the same for a
  \SI{5}{\second} batch of scans collected in the same region.  For ease of
  visualization, the batch occupancy grid has already been aligned with the
  map occupancy grid.  Panel (c) shows the cross-correlation between the
  batch and map occupancy grids at $\Delta \phi =$ \SI{0}{\degree}.  Given
  that no rotational or translational offset error has been applied to the
  batch, the correlation peak should appear at $(0,0)$. The offset of the
  peak in panel (c) from $(0,0)$ is the translational estimate error of the
  proposed method. Also note the increased positioning uncertainty in the
  along-track direction, and the two local correlation peaks (marked with red
  squares in panel (c)) due to the repeating periodic pattern of radar
  reflectors in the map and the batch (marked with red rectangles in panels
  (a) and (b)).}
  \label{fig:map-batch-xcorr}
\end{figure*}

Before diving into the quantitative analysis, it is interesting to inspect the
example of a radar batch update shown in Fig.~\ref{fig:map-batch-xcorr}.  For
ease of visualization, the batch point cloud to be localized has already been
adjusted for any translational or rotational offset from the ground truth. The
occupancy grid estimated from the \SI{5}{\second} batch of scans is shown in
Fig.~\ref{fig:batch-ogm}. Similarly, Fig.~\ref{fig:map-ogm} shows the occupancy
grid estimated from the map point cloud retrieved from the map database.
Fig.~\ref{fig:xcorr} shows the cross-correlation between the batch and map
occupancy grids. Given that the batch is already aligned with ground truth, one
should expect the correlation peak to appear at $(0,0)$ in
Fig.~\ref{fig:xcorr}. The offset of the peak from $(0,0)$ in this case would be
the translational estimate error.

Two interesting features of the cross-correlation in Fig.~\ref{fig:xcorr} are
worth noting. First, the correlation peak decays slower in the along-track
direction---in this case approximately aligned with the south-southwest
direction. This is a general feature observed throughout the dataset, since
most of the radar reflectors are aligned along the sides of the streets.
Second, there emerge two local correlation peaks offset by
$\approx$\SI{4}{\meter} along the direction of travel. These local peaks are
due to the repeating periodic structure of radar reflectors in both the map and
the batch occupancy grids. In other words, shifting the batch occupancy grid
forward or backward along the vehicle trajectory by $\approx$\SI{4}{\meter}
aligns the periodically-repeating reflectors in an off-by-one manner, leading
to another plausible solution.  Importantly, the uncertainty envelope of the
initial position estimate can span several meters, encompassing both the global
optimum and one or more local optima.  This explains why gradient-based
methods, which seek the nearest optimum, are poorly suited for use in the urban
automotive radar environment.

\subsubsection{Performance with \SI{4}{\second} Radar Batches}

\begin{figure}[ht]
  \centering
  \includegraphics[width=\linewidth]{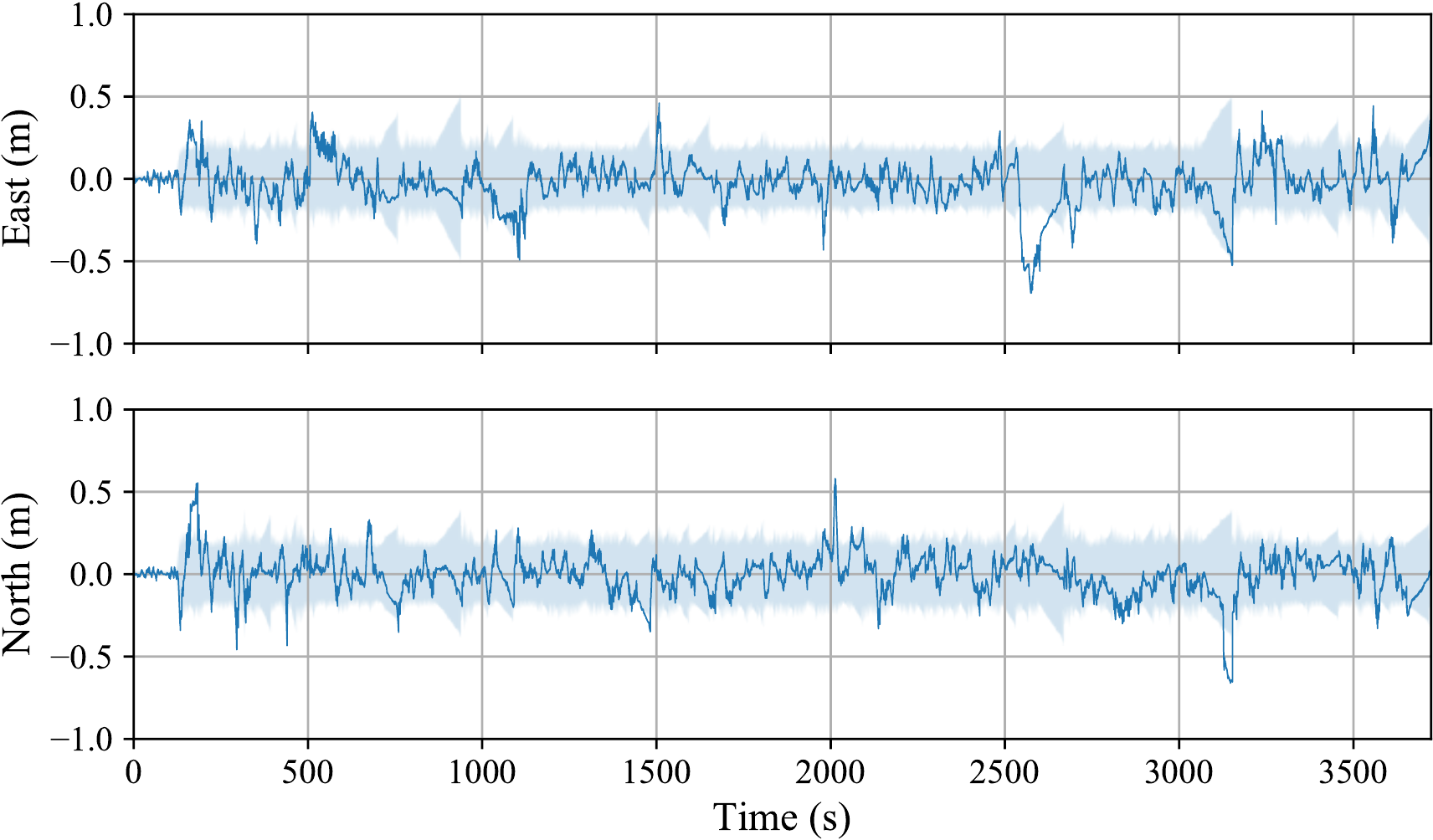}
  \caption{East and north position error time histories from field evaluation.
    In the first \SI{125}{\second} of clear-sky conditions with CDGNSS
    availability, the east and north position errors with respect to the ground
    truth are sub-decimeter, as expected. Over the subsequent \SI{60}{\minute}
    of driving in and around the urban center of the city, the proposed method
    maintains sub-35-cm (95\%) horizontal position errors. The horizontal
    position estimation errors are consistent with the predicted standard
    deviation from the EKF.}
  \label{fig:err-position}
\end{figure}

\begin{figure}[ht]
  \centering
  \includegraphics[width=\linewidth]{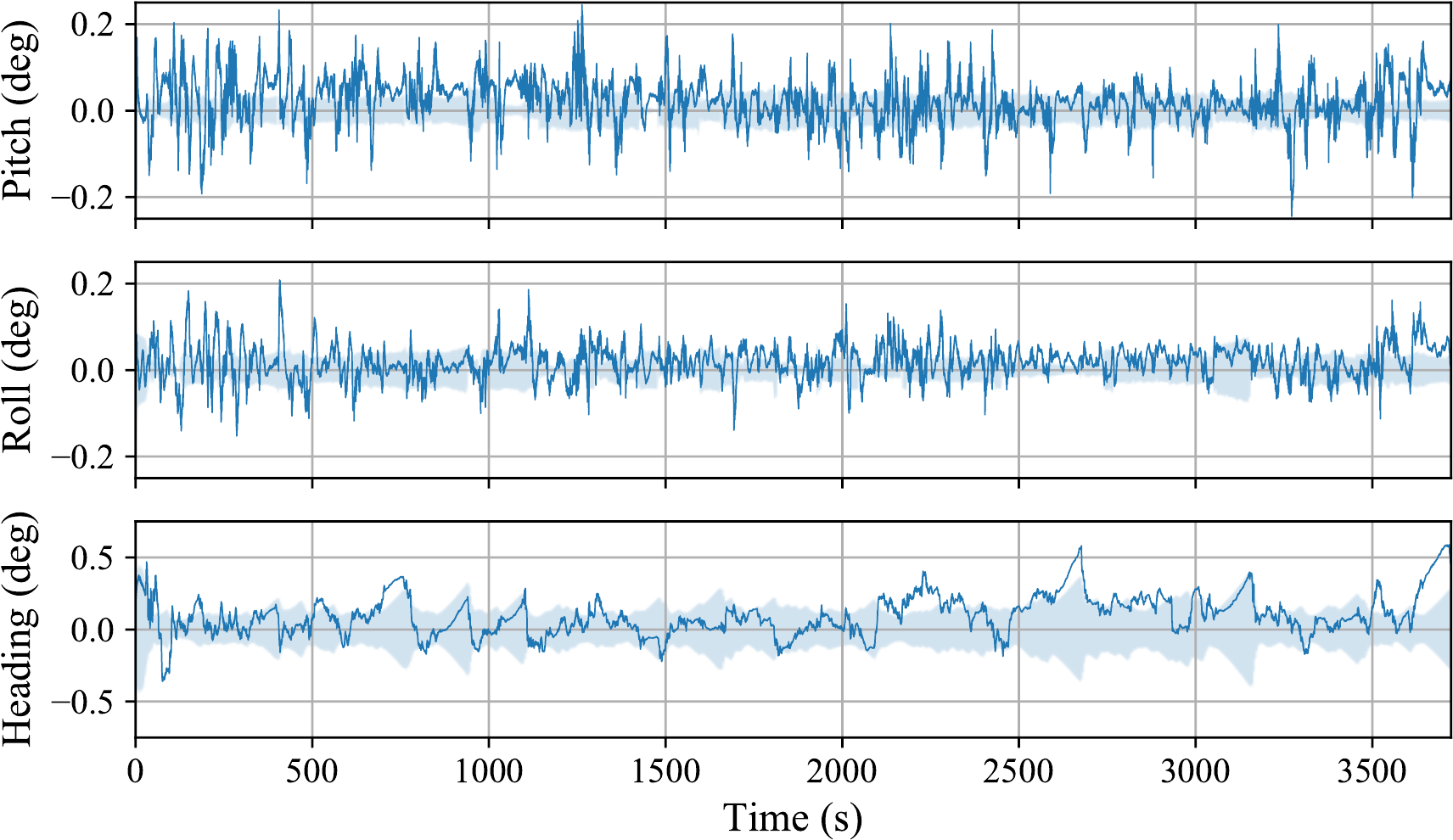}
  \caption{Vehicle orientation estimation errors from field evaluation.  The
  proposed technique maintains vehicle heading estimates to within
  \SI{0.5}{\degree} of the ground truth throughout most of the dataset, and the
  errors are consistent with the predicted uncertainty. Roll and pitch
  estimation errors are smaller and stay within \SI{0.2}{\degree} of the ground
  truth.}
  \label{fig:err-orientation}
\end{figure}

Fig.~\ref{fig:err-position} shows the east and north position error time
histories from the test scenario described above. For the results presented in
Fig.~\ref{fig:err-position} and~\ref{fig:err-orientation}, a \SI{4}{\second}
radar batch duration is chosen. In the first \SI{125}{\second} of clear-sky
conditions with CDGNSS availability, the east and north position errors with
respect to the ground truth are sub-decimeter, as expected. Over the subsequent
\SI{60}{\minute} of driving in and around the urban center of the city, the
proposed method maintains sub-35-cm horizontal position errors (95\%). The
horizontal position estimation errors are consistent with the predicted
standard deviation from the EKF. This is a remarkable result which shows that,
given a prior radar map, lane-level-accurate horizontal positioning is
achievable under zero-visibility GNSS-denied conditions with the types of
sensors that are already available on commercial vehicles. Vertical position
errors are not shown in Fig.~\ref{fig:err-position} since these are not
constrained by the two-dimensional radar batch correlation update. For ground
vehicle applications, a digital elevation map can effectively constrain errors
in altitude, if necessary.

Vehicle orientation estimation errors for the same scenario are shown in
Fig.~\ref{fig:err-orientation}. Heading estimation error, shown in the bottom
panel, is most important for ground vehicle applications. The
proposed technique maintains vehicle heading estimates to within
\SI{0.5}{\degree} of the ground truth throughout most of the dataset, and the
errors are consistent with the predicted uncertainty. Roll and pitch estimation
errors are smaller and stay within \SI{0.2}{\degree} of the ground truth.
Better estimation of roll and pitch is expected since these are directly
observable with the accelerometer measurements. The same phenomenon explains
the substantially shorter decorrelation times for roll and pitch errors as
compared to the heading error. Finally, it is noted that the EKF is mildly
inconsistent in regards to roll and pitch estimation errors. This suggests that
the accelerometer white noise and bias stability characteristics claimed in the
IMU datasheet~\cite{lordmicrostrain_3dm_gx5_25} may be optimistic in field application.

\subsubsection{Choosing a Radar Batch Length}

\begin{figure}[ht]
  \centering
  \includegraphics[width=\linewidth] {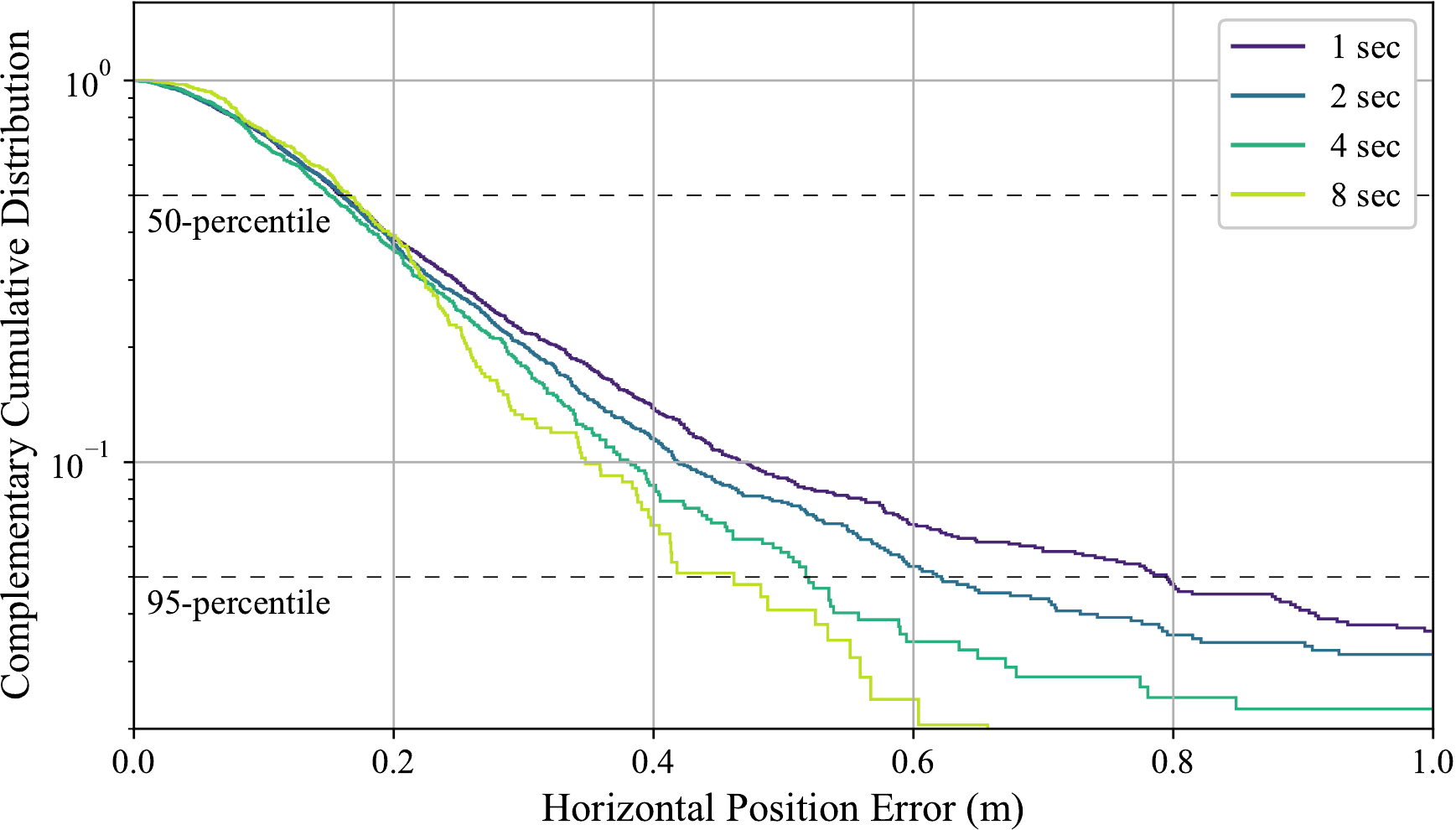}
  \caption{CCDFs for different batch lengths between \SI{1}{\second} and
  \SI{8}{\second}. The \num{50}-percentile errors are similar for shorter and
  longer batch lengths, but the difference becomes more noticeable at higher
  percentiles.}
  \label{fig:ccdf-batch-fraglen}
\end{figure}

The problem of choosing the duration of a radar batch during localization
presents an interesting trade-off. On the one hand, longer batch durations are
preferable because, intuitively, cross-correlation using a larger \emph{patch}
of the radar environment is more likely to produce a strong and unambiguous
correlation peak. Fig.~\ref{fig:ccdf-batch-fraglen} shows results from an
empirical test of this intuition. In this test, radar batches of different
durations between \SI{1}{\second} and \SI{8}{\second} were generated with
ground-truth odometry and correlated against a prior map to obtain the
estimated offset from the ground-truth pose. The complementary cumulative
distribution function (CCDF) of the horizontal position estimation errors is
shown in~\ref{fig:ccdf-batch-fraglen}. It is interesting to note that up to the
70$^{\rm th}$ percentile, errors are similar for different batch lengths. The
difference between the CCDFs becomes more pronounced at higher percentiles,
implying that errors for shorter batch lengths have heavy tails. Recall that in
the overall localization pipeline of Fig.~\ref{fig:fusion-arch}, these errors
will act as measurement errors in $\widehat{\bm{\Theta}}$. An EKF models
measurement errors to be Gaussian, which is not a good model for heavy-tailed
distributions. Accordingly, longer batch durations would appear preferable.

On the other hand, longer batches have several disadvantages. First, longer
durations between batch measurement updates leads to larger odometric drift
\emph{between} updates, as well as poorer reconstruction of the radar batch
itself. Second, some of the worst outliers due to shorter batch lengths may be
rejected in the EKF based on the $\chi^2$ NIS test, thus blunting the relative
advantage of longer batches. Shorter batch lengths allow for a larger number of
measurement updates to be performed per unit time, even if a few of those
measurements may have to be rejected as outliers.

\begin{figure}[ht]
  \centering
  \includegraphics[width=\linewidth] {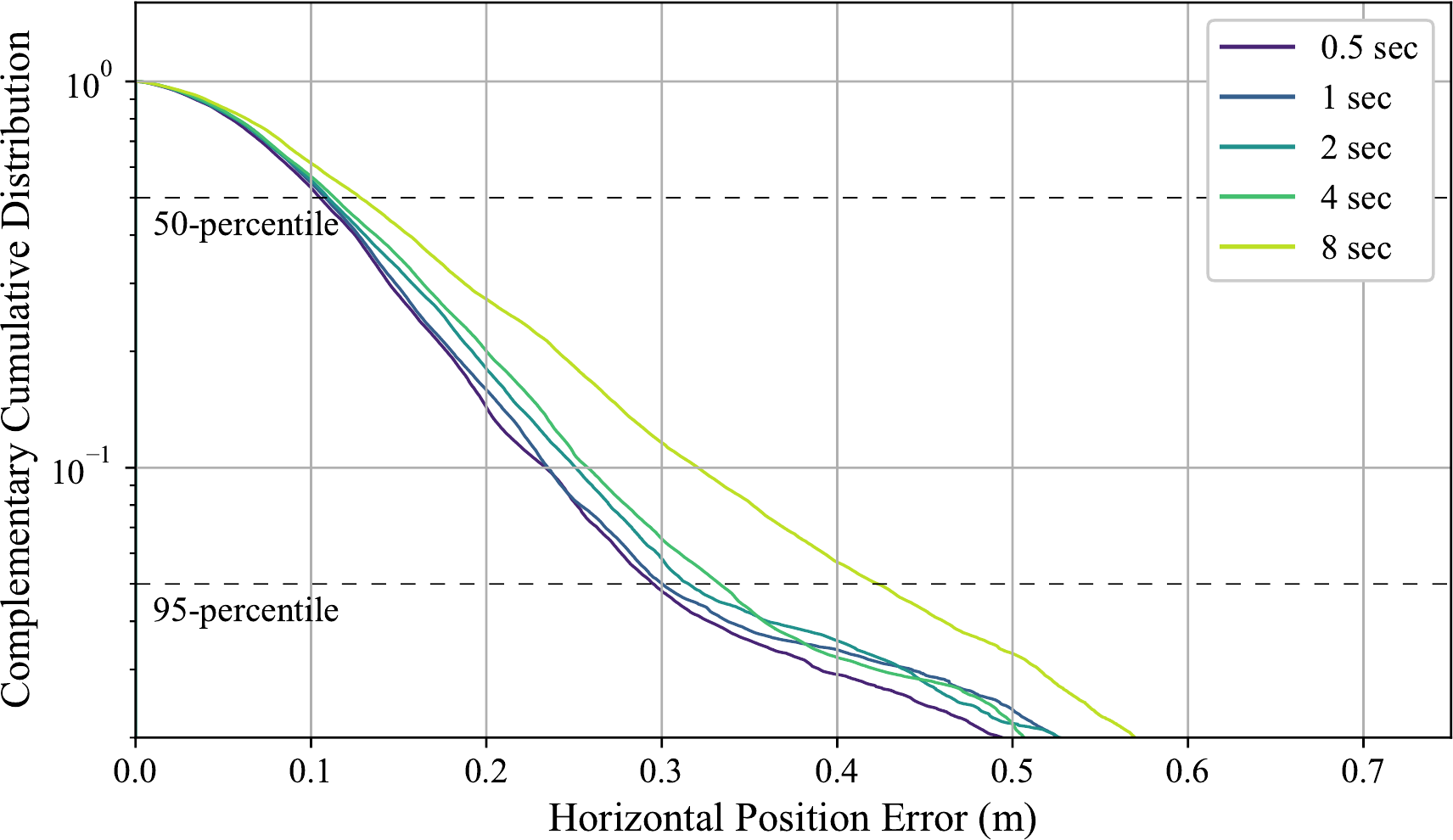}
  \caption{End-to-end effect of different batch lengths on horizontal
  positioning performance.  Other than the longest batch length of
  \SI{8}{\second}, most batch lengths appear to perform similarly well, with
  $95^{\rm th}$-percentile horizontal position errors near \SI{30}{\centi\meter}.}
  \label{fig:ccdf-position-fraglen}
\end{figure}

Fig.~\ref{fig:ccdf-position-fraglen} reveals the end-to-end effect of different
batch lengths. For a given batch length, its measurement error standard
deviation was obtained from the corresponding CCDF in
Fig.~\ref{fig:ccdf-batch-fraglen}, i.e., the $\widehat{\bm{\Theta}}$
measurement standard deviation is smaller for longer batches. Interestingly,
other than the longest batch length of \SI{8}{\second}, most batch lengths
appear to perform similarly well, with $95^{\rm}$-percentile horizontal
position errors near \SI{30}{\centi\meter}. Given the heavy-tailed nature of
measurement noise distributions when working with very short batches (from
Fig.~\ref{fig:ccdf-batch-fraglen}), batch lengths from \num{2} to
\SI{4}{\second} may be taken to be a good compromise.

\section{Conclusion}
\label{sec:conclusion}

A robust pipeline for all-weather sub-50-cm urban ground vehicle positioning
has been proposed and evaluated.  The positioning engine is based on
commercially-available low-cost automotive radars, MEMS IMU, ground vehicle
dynamics constraints, and, when available, precise GNSS measurements.
Remarkably, it has been shown that given a prior radar map, lane-level-accurate
horizontal positioning is achievable under zero-visibility GNSS-denied
conditions with the types of sensors that are already available on commercial
vehicles. In comparison with a post-processed ground truth trajectory, it was
shown that during \SI{60}{\minute} of GNSS-denied driving in the urban center
of Austin, TX, the proposed pipeline has $95^{\rm th}$-percentile errors of
\SI{35}{\centi\meter} in horizontal position and \ang{0.5} in heading.  This is
a significant development in the field of AGV localization, which has
traditionally been based on sensors such as lidar and cameras that perform
poorly in bad weather conditions.

\section*{Acknowledgments}
This work has been supported by Honda R\&D Americas through The University of
Texas Situation-Aware Vehicular Engineering Systems (SAVES) Center
(http://utsaves.org/), an initiative of the UT Wireless Networking and
Communications Group; by the National Science Foundation under Grant
No. 1454474 (CAREER); by the Data-supported Transportation Operations and
Planning Center (DSTOP), a Tier 1 USDOT University Transportation Center; and
by Sandia National Laboratories.


\begin{appendices}

\section{Partial Derivatives}
\label{app:jacobians}

\subsection{Linearized Forward Dynamics}

A few block components of $F_k$ and $G_k$ from~(\ref{eq:Fk}) and~(\ref{eq:Gk})
are listed below.
\begin{align*}
  \frac{\partial \delta\bm{p}_{k+1}^{\brm{n}}}{\partial \bm{\eta}_k^{\brm{n}}} \Bigr|_{\substack{\delta\bm{x}_k=0 \\ \bm{w}_k=0}}
  &= -\frac{T^2}{2} \left( \tilde{R}_k^{\brm{nb}} {\left[ \bm{z}_{a,k}^{\brm{b}} - \tilde{\bm{b}}_{a,k}^{\brm{b}} \right]}_\times \right) \\
  \frac{\partial \delta\bm{p}_{k+1}^{\brm{n}}}{\partial \delta\bm{b}_{a,k}^{\brm{b}}} \Bigr|_{\substack{\delta\bm{x}_k=0 \\ \bm{w}_k=0}}
  &= -\frac{T^2}{2} \tilde{R}_k^{\brm{nb}}
\end{align*}

The partial derivates of $\delta\bm{v}_{k+1}^{\brm{n}}$ with respect to
$\delta\bm{x}_k$ follow similarly.
\begin{align*}
  \frac{\partial \bm{\eta}_{k+1}^{\brm{n}}}{\partial \bm{\eta}_k^{\brm{n}}} \Bigr|_{\substack{\delta\bm{x}_k=0 \\ \bm{w}_k=0}}
  &\approx I_{3 \times 3} \\
  \frac{\partial \bm{\eta}_{k+1}^{\brm{n}}}{\partial \delta\bm{b}_{\omega,k}^{\brm{b}}} \Bigr|_{\substack{\delta\bm{x}_k=0 \\ \bm{w}_k=0}}
  &\approx -T \tilde{R}_{k+1}^{\brm{nb}} \mathrm{J}_r \left( \frac{T}{2} \left( \bm{z}_{\omega,k}^{\brm{b}} - \tilde{\bm{b}}_{\omega,k}^{\brm{b}} - \tilde{R}_k^{\brm{bn}} \bm{\omega}_{\brm{e}}^{\brm{n}} \right) \right) \\
  &\approx -T \tilde{R}_{k+1}^{\brm{nb}}
\end{align*}
where
\begin{equation*}
  \mathrm{J}_r \left( \bm{\theta} \right) = I_{3 \times 3} - \frac{1 - \cos{\| \bm{\theta} \|}}{{\| \bm{\theta} \|}^2} {\left[ \bm{\theta} \right]}_\times + \frac{\| \bm{\theta} \| - \sin{\| \bm{\theta} \|}}{{\| \bm{\theta} \|}^3} {\left[ \bm{\theta} \right]}_\times^2
\end{equation*}
is the right Jacobian of $SO(3)$~\cite{sola2017quaternion}.
\begin{align*}
  \frac{\partial \bm{\eta}_{k+1}^{\brm{n}}}{\partial \bm{w}_{\omega,k}} \Bigr|_{\substack{\delta\bm{x}_k=0 \\ \bm{w}_k=0}}
  &\approx T \tilde{R}_{k+1}^{\brm{nb}} \mathrm{J}_r \left( \frac{T}{2} \left( \bm{z}_{\omega,k}^{\brm{b}} - \tilde{\bm{b}}_{\omega,k}^{\brm{b}} - \tilde{R}_k^{\brm{bn}} \bm{\omega}_{\brm{e}}^{\brm{n}} \right) \right) \\
  &\approx T \tilde{R}_{k+1}^{\brm{nb}}
\end{align*}

\subsection{Linearized Measurement Models}

The partial derivative of the measurement $\bm{z}_{\brm{a}_i,k}^{\brm{n}}$
from~(\ref{eq:cdgnss}) can be expressed as
\begin{equation*}
  \frac{\partial \bm{z}_{\brm{a}_i, k}^{\brm{n}}}{\partial \delta\bm{x}_k} \Bigr|_{\substack{\delta\bm{x}_k=0 \\ \bm{e}_{\brm{a}_i, k}=0}}
  = \frac{\partial \bm{z}_{\brm{a}_i, k}^{\brm{n}}}{\partial \bm{x}_k} \Bigr|_{\substack{\delta\bm{x}_k=0 \\ \bm{e}_{\brm{a}_i, k}=0}} \cdot
    \frac{\partial \bm{x}_k}{\partial \delta\bm{x}_k} \Bigr|_{\substack{\delta\bm{x}_k=0 \\ \bm{e}_{\brm{a}_i, k}=0}}
\end{equation*}
where the non-trivial block matrices are as follows:
\begin{align*}
  \frac{\partial \bm{z}_{\brm{a}_i, k}^{\brm{n}}}{\partial \bm{q}_k^{\brm{nb}}} \Bigr|_{\substack{\delta\bm{x}_k=0 \\ \bm{e}_{\brm{a}_i, k}=0}}
  &= \frac{\partial \left( \bm{q}_k^{\brm{nb}} \odot R^{\brm{bs}} \bm{p}_{\brm{ba}_i}^{\brm{s}} \odot \bm{q}_k^{\brm{bn}} \right)}{\partial \bm{q}_k^{\brm{nb}}} \\
  \frac{\partial \bm{q}_k^{\brm{nb}}}{\partial \bm{\eta}_k^{\brm{n}}} \Bigr|_{\substack{\delta\bm{x}_k=0 \\ \bm{e}_{\brm{a}_i, k}=0}}
  &= \frac{1}{2} \begin{bmatrix} -q_x & -q_y & -q_z \\ q_w & q_z & -q_y \\ -q_z & q_w & q_x \\ q_y & -q_x & q_w \end{bmatrix}
\end{align*}
with $\tilde{\bm{q}}_k^{\brm{nb}} = \left[ q_w, q_x, q_y, q_z \right]$. The
expression for derivative of the rotation with respect to the quaternion can
be found in~\cite[Sec.  4.3.2]{sola2017quaternion}.

For the radar range rate measurement $\bm{z}_{\brm{r}_i,k}^{\brm{r}_i}$
\begin{align*}
  \frac{\partial \bm{z}_{\brm{r}_i,k}^{\brm{r}_i}}{\partial \bm{v}_k^{\brm{n}}}
  &= R^{\brm{r}_i \brm{s}} R^{\brm{sb}} \tilde{R}_k^{\brm{bn}} \\
  \frac{\partial \bm{z}_{\brm{r}_i,k}^{\brm{r}_i}}{\partial \bm{q}_k^{\brm{nb}}}
  &= R^{\brm{r}_i \brm{s}} R^{\brm{sb}} \frac{\partial \left( \bm{q}_k^{\brm{bn}} \odot \tilde{\bm{v}}_k^{\brm{n}} \odot \bm{q}_k^{\brm{nb}} \right)}{\partial \bm{q}_k^{\brm{nb}}} \\
  \frac{\partial \bm{z}_{\brm{r}_i,k}^{\brm{r}_i}}{\partial \bm{b}_{\omega,k}^{\brm{b}}}
  &= -R^{\brm{r}_i \brm{s}} R^{\brm{sb}} {\left[ R^{\brm{bs}} \bm{p}_{\brm{br}_i}^{\brm{s}} \right]}_\times
\end{align*}

The partial derivatives of $\bm{z}_{\mathrm{nhc},k}^{\brm{v}}$ and
$\bm{z}_{\mathrm{zupt},k}^{\brm{v}}$ follow similarly.

\section{Nonlinear Error-State Rauch-Tung-Striebel Smoother}
\label{app:smoothing}

The conventional expression for the extended Rauch-Tung-Striebel (RTS) smoother
is given as~\cite[Chap. 9]{sarkka2013bayesian}
\begin{align*}
  \bm{x}_k^\star &= \hat{\bm{x}}_k + C_k \left( \bm{x}_{k+1}^\star - f_k \left( \hat{\bm{x}}_k \right) \right) \\
  P_k^\star &= P_k + C_k \left( P_{k+1}^\star - F_k P_k F_k^\top - G_k Q_k G_k^\top \right) C_k^\top
\end{align*}
with
\begin{equation*}
  C_k = P_k F_k^\top {\left( F_k P_k F_k^\top + G_k Q_k G_k^\top \right)}^{-1}
\end{equation*}
where $^\star$ indicates the smoothed estimate and $\hat{}$ indicates the
filtered estimate. This expression is derived by linearizing the dynamics
at the \emph{filtered} state estimate during the \emph{backward smoothing}
pass.

In contrast, this paper prefers to linearize the dynamics at the predicted
smoothed estimate $\bar{\bm{x}}_k^\star$ instead
\begin{equation*}
  \bar{\bm{x}}_k^\star \triangleq f_k^{-1} \left( \bm{x}_{k+1}^\star, \bm{u}_k, \bm{0} \right)
\end{equation*}

This formulation results in a similar but slightly modified expression for the
extended RTS smoother
\begin{align*}
  \bm{x}_k^\star &= \hat{\bm{x}}_k + C_k^\star F_k^\star \left( \bar{\bm{x}}_k^\star - \hat{\bm{x}}_k \right) \\
  P_k^\star &= P_k + C_k^\star \left( P_{k+1}^\star - F_k^\star P_k F_k^{\star\top} - G_k^\star Q_k G_k^{\star\top} \right) C_k^{\star\top}
\end{align*}
with
\begin{equation*}
  C_k^\star = P_k F_k^{\star\top} {\left( F_k^\star P_k F_k^{\star\top} + G_k^\star Q_k G_k^{\star\top} \right)}^{-1}
\end{equation*}
where $F_k^\star$ and $G_k^\star$ denote linearized forward dynamics around
$\bar{\bm{x}}_k^\star$.


\end{appendices}

\bibliographystyle{IEEEtran}
\bibliography{pangea}

\begin{thebibliography}{10}
\providecommand{\url}[1]{#1}
\csname url@samestyle\endcsname
\providecommand{\newblock}{\relax}
\providecommand{\bibinfo}[2]{#2}
\providecommand{\BIBentrySTDinterwordspacing}{\spaceskip=0pt\relax}
\providecommand{\BIBentryALTinterwordstretchfactor}{4}
\providecommand{\BIBentryALTinterwordspacing}{\spaceskip=\fontdimen2\font plus
\BIBentryALTinterwordstretchfactor\fontdimen3\font minus
  \fontdimen4\font\relax}
\providecommand{\BIBforeignlanguage}[2]{{%
\expandafter\ifx\csname l@#1\endcsname\relax
\typeout{** WARNING: IEEEtran.bst: No hyphenation pattern has been}%
\typeout{** loaded for the language `#1'. Using the pattern for}%
\typeout{** the default language instead.}%
\else
\language=\csname l@#1\endcsname
\fi
#2}}
\providecommand{\BIBdecl}{\relax}
\BIBdecl

\bibitem{fajardo2011automated}
D.~Fajardo, T.-C. Au, S.~Waller, P.~Stone, and D.~Yang, ``Automated
  intersection control: Performance of future innovation versus current traffic
  signal control,'' \emph{Transportation Research Record: Journal of the
  Transportation Research Board}, no. 2259, pp. 223--232, 2011.

\bibitem{choi2016mmWaveVehicular}
J.~Choi, V.~Va, N.~Gonzalez-Prelcic, R.~Daniels, C.~R. Bhat, and R.~W. Heath,
  ``Millimeter-wave vehicular communication to support massive automotive
  sensing,'' \emph{IEEE Communications Magazine}, vol.~54, no.~12, pp.
  160--167, December 2016.

\bibitem{lachapelle2020riskIcassp}
D.~La{C}hapelle, T.~E. Humphreys, L.~Narula, P.~A. Iannucci, and
  E.~Moradi-Pari, ``Automotive collision risk estimation under cooperative
  sensing,'' in \emph{Proceedings of the IEEE International Conference on
  Acoustics, Speech, and Signal Processing}, Barcelona, Spain, 2020.

\bibitem{yen2015evaluation}
K.~S. Yen, C.~Shankwitz, B.~Newstrom, T.~A. Lasky, and B.~Ravani, ``Evaluation
  of the {U}niversity of {M}innesota {GPS} {S}nowplow {D}river {A}ssistance
  {P}rogram,'' California Department of Transportation, Tech. Rep., 2015.

\bibitem{petovello2004benefits}
M.~Petovello, M.~Cannon, and G.~Lachapelle, ``Benefits of using a
  tactical-grade {IMU} for high-accuracy positioning,'' \emph{Navigation,
  Journal of the Institute of Navigation}, vol.~51, no.~1, pp. 1--12, 2004.

\bibitem{scherzinger2006precise}
B.~M. Scherzinger, ``Precise robust positioning with inertially aided {RTK},''
  \emph{Navigation}, vol.~53, no.~2, pp. 73--83, 2006.

\bibitem{zhangComparisonWithTactical2006}
H.~T. Zhang, ``Performance comparison on kinematic {GPS} integrated with
  different tactical-grade {IMUs},'' Master's thesis, The University of
  Calgary, Jan. 2006.

\bibitem{kennedy2006architecture}
S.~Kennedy, J.~Hamilton, and H.~Martell, ``Architecture and system performance
  of {SPAN}---{NovAtel's} {GPS/INS} solution,'' in \emph{Position, Location,
  And Navigation Symposium, 2006 IEEE/ION}.\hskip 1em plus 0.5em minus
  0.4em\relax IEEE, 2006, p. 266.

\bibitem{humphreys2019deepUrbanIts}
T.~E. Humphreys, M.~J. Murrian, and L.~Narula, ``Deep urban unaided precise
  {Global Navigation Satellite System} vehicle positioning,'' \emph{IEEE
  Intelligent Transportation Systems Magazine}, 2020.

\bibitem{humphreysGNSShandbook}
T.~E. Humphreys, \emph{Springer Handbook of Global Navigation Satellite
  Systems}.\hskip 1em plus 0.5em minus 0.4em\relax Springer, 2017, ch.
  Interference, pp. 469--504.

\bibitem{narula2018accurate}
\BIBentryALTinterwordspacing
L.~Narula, J.~M. Wooten, M.~J. Murrian, D.~M. LaChapelle, and T.~E. Humphreys,
  ``Accurate collaborative globally-referenced digital mapping with standard
  {GNSS},'' \emph{Sensors}, vol.~18, no.~8, 2018. [Online]. Available:
  \url{http://www.mdpi.com/1424-8220/18/8/2452}
\BIBentrySTDinterwordspacing

\bibitem{narula2020texcup}
L.~Narula, D.~M. LaChapelle, M.~J. Murrian, J.~M. Wooten, T.~E. Humphreys,
  J.-B. Lacambre, E.~de~Toldi, and G.~Morvant, ``{TEX-CUP}: {T}he {U}niversity
  of {T}exas {C}hallenge for {U}rban {P}ositioning,'' in \emph{Proceedings of
  the IEEE/ION PLANSx Meeting}, 2020.

\bibitem{narula2020radarpositioning}
L.~Narula, P.~A. Iannucci, and T.~E. Humphreys, ``Automotive-radar-based 50-cm
  urban positioning,'' in \emph{Proceedings of the IEEE/ION PLANSx Meeting},
  2020.

\bibitem{chetverikov2002trimmed}
D.~Chetverikov, D.~Svirko, D.~Stepanov, and P.~Krsek, ``The trimmed iterative
  closest point algorithm,'' in \emph{Object recognition supported by user
  interaction for service robots}, vol.~3.\hskip 1em plus 0.5em minus
  0.4em\relax IEEE, 2002, pp. 545--548.

\bibitem{ward2016vehicle}
E.~Ward and J.~Folkesson, ``Vehicle localization with low cost radar sensors,''
  in \emph{2016 IEEE Intelligent Vehicles Symposium (IV)}.\hskip 1em plus 0.5em
  minus 0.4em\relax IEEE, 2016, pp. 864--870.

\bibitem{holder2019real}
M.~Holder, S.~Hellwig, and H.~Winner, ``Real-time pose graph {SLAM} based on
  radar,'' in \emph{2019 IEEE Intelligent Vehicles Symposium (IV)}.\hskip 1em
  plus 0.5em minus 0.4em\relax IEEE, 2019, pp. 1145--1151.

\bibitem{tsin2004correlation}
Y.~Tsin and T.~Kanade, ``A correlation-based approach to robust point set
  registration,'' in \emph{European conference on computer vision}.\hskip 1em
  plus 0.5em minus 0.4em\relax Springer, 2004, pp. 558--569.

\bibitem{jian2010robust}
B.~Jian and B.~C. Vemuri, ``Robust point set registration using {G}aussian
  mixture models,'' \emph{IEEE Transactions on Pattern Analysis and Machine
  Intelligence}, vol.~33, no.~8, pp. 1633--1645, 2010.

\bibitem{myronenko2010point}
A.~Myronenko and X.~Song, ``Point set registration: {C}oherent point drift,''
  \emph{IEEE Transactions on Pattern Analysis and Machine Intelligence},
  vol.~32, no.~12, pp. 2262--2275, 2010.

\bibitem{gao2019filterreg}
W.~Gao and R.~Tedrake, ``{F}ilter{R}eg: {R}obust and efficient probabilistic
  point-set registration using gaussian filter and twist parameterization,'' in
  \emph{Proceedings of the IEEE Conference on Computer Vision and Pattern
  Recognition}, 2019, pp. 11\,095--11\,104.

\bibitem{cen2018precise}
S.~H. Cen and P.~Newman, ``Precise ego-motion estimation with millimeter-wave
  radar under diverse and challenging conditions,'' in \emph{2018 IEEE
  International Conference on Robotics and Automation (ICRA)}.\hskip 1em plus
  0.5em minus 0.4em\relax IEEE, 2018, pp. 1--8.

\bibitem{cen2019radar}
------, ``Radar-only ego-motion estimation in difficult settings via graph
  matching,'' \emph{arXiv preprint arXiv:1904.11476}, 2019.

\bibitem{barnes2020under}
D.~Barnes and I.~Posner, ``Under the radar: {L}earning to predict robust
  keypoints for odometry estimation and metric localisation in radar,''
  \emph{arXiv preprint arXiv:2001.10789}, 2020.

\bibitem{barnes2019masking}
D.~Barnes, R.~Weston, and I.~Posner, ``Masking by moving: {L}earning
  distraction-free radar odometry from pose information,'' \emph{arXiv preprint
  arXiv:1909.03752}, 2019.

\bibitem{schuster2016landmark}
F.~Schuster, C.~G. Keller, M.~Rapp, M.~Haueis, and C.~Curio, ``Landmark based
  radar {SLAM} using graph optimization,'' in \emph{Intelligent Transportation
  Systems (ITSC), 2016 IEEE 19th International Conference on}.\hskip 1em plus
  0.5em minus 0.4em\relax IEEE, 2016, pp. 2559--2564.

\bibitem{schoen2017real}
M.~Schoen, M.~Horn, M.~Hahn, and J.~Dickmann, ``Real-time radar {SLAM}.''

\bibitem{callmer2011radar}
J.~Callmer, D.~T{\"o}rnqvist, F.~Gustafsson, H.~Svensson, and P.~Carlbom,
  ``Radar {SLAM} using visual features,'' \emph{EURASIP Journal on Advances in
  Signal Processing}, vol. 2011, no.~1, p.~71, 2011.

\bibitem{hong2020radarslam}
Z.~Hong, Y.~Petillot, and S.~Wang, ``Radar{SLAM}: {R}adar based large-scale
  {SLAM} in all weathers,'' \emph{arXiv preprint arXiv:2005.02198}, 2020.

\bibitem{yoneda2018vehicle}
K.~Yoneda, N.~Hashimoto, R.~Yanase, M.~Aldibaja, and N.~Suganuma, ``Vehicle
  localization using 76{GHz} omnidirectional millimeter-wave radar for winter
  automated driving,'' in \emph{2018 IEEE Intelligent Vehicles Symposium
  (IV)}.\hskip 1em plus 0.5em minus 0.4em\relax IEEE, 2018, pp. 971--977.

\bibitem{mahler2003multitargetFirstOrder}
R.~P. Mahler, ``Multitarget {B}ayes filtering via first-order multitarget
  moments,'' \emph{IEEE Transactions on Aerospace and Electronic systems},
  vol.~39, no.~4, pp. 1152--1178, 2003.

\bibitem{erdinc2009bin}
O.~Erdinc, P.~Willett, and Y.~Bar-Shalom, ``The bin-occupancy filter and its
  connection to the {PHD} filters,'' \emph{IEEE Transactions on Signal
  Processing}, vol.~57, no.~11, pp. 4232--4246, 2009.

\bibitem{mullane2011random}
J.~Mullane, B.-N. Vo, M.~D. Adams, and B.-T. Vo, ``A random-finite-set approach
  to {B}ayesian {SLAM},'' \emph{IEEE Transactions on Robotics}, vol.~27, no.~2,
  pp. 268--282, 2011.

\bibitem{deusch2015labeled}
H.~Deusch, S.~Reuter, and K.~Dietmayer, ``The labeled multi-{B}ernoulli {SLAM}
  filter,'' \emph{IEEE Signal Processing Letters}, vol.~22, no.~10, pp.
  1561--1565, 2015.

\bibitem{stubler2017continuously}
M.~St{\"u}bler, S.~Reuter, and K.~Dietmayer, ``A continuously learning
  feature-based map using a {B}ernoulli filtering approach,'' in \emph{2017
  Sensor Data Fusion: Trends, Solutions, Applications (SDF)}.\hskip 1em plus
  0.5em minus 0.4em\relax IEEE, 2017, pp. 1--6.

\bibitem{fatemi2017poisson}
M.~Fatemi, K.~Granstr{\"o}m, L.~Svensson, F.~J. Ruiz, and L.~Hammarstrand,
  ``Poisson multi-bernoulli mapping using {G}ibbs sampling,'' \emph{IEEE
  Transactions on Signal Processing}, vol.~65, no.~11, pp. 2814--2827, 2017.

\bibitem{qin2018vins}
T.~Qin, P.~Li, and S.~Shen, ``{VINS}-mono: A robust and versatile monocular
  visual-inertial state estimator,'' \emph{IEEE Transactions on Robotics},
  vol.~34, no.~4, pp. 1004--1020, 2018.

\bibitem{mur2017visual}
R.~Mur-Artal and J.~D. Tard{\'o}s, ``Visual-inertial monocular {SLAM} with map
  reuse,'' \emph{IEEE Robotics and Automation Letters}, vol.~2, no.~2, pp.
  796--803, 2017.

\bibitem{chiang2020navigation}
K.-W. Chiang, G.-J. Tsai, Y.-H. Li, Y.~Li, and N.~El-Sheimy, ``Navigation
  engine design for automated driving using {INS/GNSS/3D LiDAR-SLAM} and
  integrity assessment,'' \emph{Remote Sensing}, vol.~12, no.~10, p. 1564,
  2020.

\bibitem{forster2013collaborative}
C.~Forster, S.~Lynen, L.~Kneip, and D.~Scaramuzza, ``Collaborative monocular
  slam with multiple micro aerial vehicles,'' in \emph{2013 IEEE/RSJ
  International Conference on Intelligent Robots and Systems}, Nov 2013, pp.
  3962--3970.

\bibitem{steder2008visual}
B.~Steder, G.~Grisetti, C.~Stachniss, and W.~Burgard, ``Visual {SLAM} for
  flying vehicles,'' \emph{IEEE Transactions on Robotics}, vol.~24, no.~5, pp.
  1088--1093, 2008.

\bibitem{ye2019tightly}
H.~Ye, Y.~Chen, and M.~Liu, ``Tightly coupled 3d lidar inertial odometry and
  mapping,'' in \emph{2019 International Conference on Robotics and Automation
  (ICRA)}.\hskip 1em plus 0.5em minus 0.4em\relax IEEE, 2019, pp. 3144--3150.

\bibitem{li2014lidar}
R.~Li, J.~Liu, L.~Zhang, and Y.~Hang, ``Lidar/mems imu integrated navigation
  (slam) method for a small uav in indoor environments,'' in \emph{2014 DGON
  Inertial Sensors and Systems (ISS)}.\hskip 1em plus 0.5em minus 0.4em\relax
  IEEE, 2014, pp. 1--15.

\bibitem{barra2019localization}
J.~Barra, S.~Lesecq, M.~Zarudniev, O.~Debicki, N.~Mareau, and L.~Ouvry,
  ``Localization system in {GPS}-denied environments using radar and {IMU}
  measurements: {A}pplication to a smart white cane,'' in \emph{2019 18th
  European Control Conference (ECC)}.\hskip 1em plus 0.5em minus 0.4em\relax
  IEEE, 2019, pp. 1201--1206.

\bibitem{almalioglu2019milli}
Y.~Almalioglu, M.~Turan, C.~X. Lu, N.~Trigoni, and A.~Markham, ``Milli-{RIO}:
  {E}go-motion estimation with low-cost millimetre-wave radar,'' \emph{arXiv
  preprint arXiv:1909.05774}, 2019.

\bibitem{kramer2020radar}
A.~Kramer, C.~Stahoviak, A.~Santamaria-Navarro, A.-A. Agha-Mohammadi, and
  C.~Heckman, ``Radar-inertial ego-velocity estimation for visually degraded
  environments,'' in \emph{2020 IEEE International Conference on Robotics and
  Automation (ICRA)}.\hskip 1em plus 0.5em minus 0.4em\relax IEEE, 2020.

\bibitem{thrun2005probabilistic}
S.~Thrun, W.~Burgard, and D.~Fox, \emph{Probabilistic robotics}.\hskip 1em plus
  0.5em minus 0.4em\relax MIT press, 2005.

\bibitem{sola2017quaternion}
J.~Sola, ``Quaternion kinematics for the error-state {K}alman filter,''
  \emph{arXiv preprint arXiv:1711.02508}, 2017.

\bibitem{kok2017inertial}
M.~Kok, J.~D. Hol, and T.~B. Sch{\"o}n, ``Using inertial sensors for position
  and orientation estimation,'' \emph{arXiv preprint arXiv:1704.06053}, 2017.

\bibitem{crassidis2007survey}
J.~L. Crassidis, F.~L. Markley, and Y.~Cheng, ``Survey of nonlinear attitude
  estimation methods,'' \emph{Journal of guidance control and dynamics},
  vol.~30, no.~1, p.~12, 2007.

\bibitem{woodman2007introduction}
O.~Woodman, ``An introduction to inertial navigation,'' \emph{University of
  Cambridge, Computer Laboratory, Tech. Rep. UCAMCL-TR-696}, 2007.

\bibitem{skog2010zero}
I.~Skog, P.~Handel, J.-O. Nilsson, and J.~Rantakokko, ``Zero-velocity
  detection--{A}n algorithm evaluation,'' \emph{IEEE transactions on biomedical
  engineering}, vol.~57, no.~11, pp. 2657--2666, 2010.

\bibitem{sarkka2013bayesian}
S.~S{\"a}rkk{\"a}, \emph{Bayesian filtering and smoothing}.\hskip 1em plus
  0.5em minus 0.4em\relax Cambridge University Press, 2013, vol.~3.

\bibitem{y_barshalom01_tan}
Y.~Bar-Shalom, X.~R. Li, and T.~Kirubarajan, \emph{Estimation with Applications
  to Tracking and Navigation}.\hskip 1em plus 0.5em minus 0.4em\relax New York:
  John Wiley and Sons, 2001.

\bibitem{narula2018accuracyLimits}
L.~Narula, M.~J. Murrian, and T.~E. Humphreys, ``Accuracy limits for
  globally-referenced digital mapping using standard {GNSS},'' in \emph{2018
  21st International Conference on Intelligent Transportation Systems
  (ITSC)}.\hskip 1em plus 0.5em minus 0.4em\relax IEEE, 2018, pp. 3075--3082.

\bibitem{lordmicrostrain_3dm_gx5_25}
{LORD Sensing MicroStrain}, ``{3DM-GX5-25 Attitude and Heading Reference
  System},'' \url{https://bit.ly/32CKIaO}, accessed 2020-08-31.

\end{thebibliography}

\balance

\end{document}